\documentclass[prb,aps,twocolumn,amsmath,amssymb,floatfix,superscriptaddress]{revtex4-2}
\usepackage{amsmath,bm}
\usepackage{latexsym}
\usepackage{amssymb}
\usepackage[colorlinks=true, citecolor=blue, urlcolor=blue ]{hyperref}
\usepackage{epsf,graphics,graphicx}
\DeclareMathAlphabet{\mathpzc}{OT1}{pzc}{m}{it}
\DeclareFontFamily{OT1}{pzc}{}
\DeclareFontShape{OT1}{pzc}{m}{it}{ <-> s*[1.1] pzcmi7t }{}
\usepackage{contour}

\begin{document}

\title{Persistent Charge and Spin Currents in a Ferromagnetic Hatano-Nelson Ring}

\author{Sourav Karmakar}

\email{karmakarsourav2015@gmail.com}

\affiliation{Physics and Applied Mathematics Unit, Indian Statistical Institute, 203 Barrackpore Trunk Road, Kolkata-700 108, India}

\author{Sudin Ganguly}

\email[corresponding author:] {sudinganguly@gmail.com}

\affiliation{Department of Physics, Adamas University, Adamas Knowledge City, Barasat-Barrackpore Road, 24 Parganas North, Kolkata-700 126, India}

\author{Santanu K. Maiti}

\email{santanu.maiti@isical.ac.in}

\affiliation{Physics and Applied Mathematics Unit, Indian Statistical Institute, 203 Barrackpore Trunk Road, Kolkata-700 108, India}

\begin{abstract}
We investigate persistent charge and spin currents in a ferromagnetic Hatano-Nelson ring with anti-Hermitian intradimer hopping, where non-reciprocal hopping generates a synthetic magnetic flux and drives a non-Hermitian Aharonov-Bohm effect. The system supports both real and imaginary persistent currents, with ferromagnetic spin splitting enabling all three spin-current components, dictated by the orientation of magnetic moments. The currents are computed using the current operator method within a biorthogonal basis. In parallel, the complex band structure is analyzed to uncover the spectral characteristics. We emphasize how the currents evolve across different topological regimes, and how they are influenced by chemical potential, ferromagnetic ordering, finite size, and disorder. Strikingly, disorder can even amplify spin currents, opening powerful new routes for manipulating spin transport in non-Hermitian systems.

\end{abstract}
\maketitle

\section{Introduction}
In mesoscopic physics, persistent current has emerged as a hallmark quantum-coherent phenomenon. It occurs in a metallic ring when the electronic mean free path exceeds the ring circumference in the presence of a magnetic flux, a concept first proposed by Kulik~\cite{pc1} and subsequently formulated by B\"{u}ttiker, Imry, and Landauer in the framework of a one-dimensional disordered ring~\cite{pc2}. Experimental confirmations, ranging from ensembles of copper rings~\cite{pcex1} to isolated gold loops~\cite{pcex2}, firmly established its existence. Since then, persistent current has been extensively studied under a variety of conditions, both theoretically~\cite{pc3,pc4,pc5,pc6,pc7,pc8,pc9,pc10,pc11} and experimentally~\cite{pcex3,pcex4,pcex5,pcex6}.

The concept of  persistent spin currents emerged in the early 1990s, when Loss and co-workers showed that Berry's phase could generate both charge and spin currents in mesoscopic rings \cite{spc90}. Soon after, the Aharonov-Casher mechanism and the role of spin-orbit coupling were established in one-dimensional rings \cite{spc94,spc95}, leading to proposals such as the spin-interference device for spintronics applications \cite{spc99}. Subsequent studies examined Rashba spin-orbit coupling, Heisenberg and ferrimagnetic spin rings, and generalized the concept to systems with impurities and angular spin currents \cite{spc03a,spc03b,spc05,spc07,spc08,spcding}. More recent works investigated spin-polarized transport in Rashba-coupled rings attached to reservoirs \cite{spc14}, while our group analyzed charge and  persistent spin currents in spin-orbit-coupled and Fibonacci rings \cite{maiti14,maiti16}. Extensions also include spin dynamics in Aharonov-Casher rings with inhomogeneous Rashba interaction and the impact of internal Zeeman fields \cite{spc21,spc23}.

As mentioned above, investigations of charge and  persistent spin currents have traditionally relied on theoretical models within the Hermitian framework. In recent years, however, the rise of non-Hermitian quantum mechanics~\cite{hatano1,hatano2}, especially in open and dissipative systems, has revealed novel transport phenomena beyond conventional Hermitian settings~\cite{nhreview}. Motivated by these advances, the study of persistent currents in non-Hermitian systems has only recently begun~\cite{li2021,ratchet2022,shen2024,ganguly2025,ghosh2025,sarkar2025}. The mechanisms by which non-Hermiticity enters these models vary significantly. For example, Refs.~\cite{li2021,ganguly2025} employed a Hatano-Nelson (HN) type model with anti-Hermitian intradimer hopping, which enables persistent currents without magnetic flux. In this case, asymmetric couplings act as a synthetic gauge field~\cite{hatano1,hatano2} that drives the current. The later work~\cite{ganguly2025} further examined the role of correlated and uncorrelated disorder, showing that disorder can amplify the persistent current. In contrast, Ref.~\cite{ratchet2022} analyzed a driven-dissipative bosonic system, where persistent flow of atoms emerges even without any dynamic drive or nonlocal dissipation. Dissipative non-Hermitian effects have also been studied in electronic systems such as a phase-biased superconducting-normal-superconducting junction and a flux-threaded metallic ring~\cite{shen2024}, where a generalized non-Hermitian Fermi-Dirac distribution provides analytical expressions for persistent current valid even at exceptional points. Additional developments include the study of a parity-time symmetric extended SSH chain with boundary-induced currents~\cite{ghosh2025}, and disordered rings described by the Aubry-Andr\'{e}-Harper potential with added non-Hermiticity, which display enhanced persistent currents~\cite{sarkar2025}.

Although some progress has been made in understanding flux-sensitive responses in non-Hermitian lattice systems, the role of spin degrees of freedom in non-Hermitian persistent transport remains largely unexplored. While Hatano and Nelson did not explicitly formulate persistent currents in the mesoscopic sense, their demonstration of strong spectral sensitivity to boundary conditions and synthetic gauge phases in non-reciprocal lattices laid the foundation for later interpretations of current-carrying responses in non-Hermitian rings~\cite{hatano1,hatano2}.

In Hermitian systems, persistent spin currents are often closely connected to their charge counterparts and can frequently be understood within a unified framework. In contrast, non-Hermitian Hamiltonians fundamentally alter the structure of eigenstates and expectation values, necessitating a biorthogonal description in which spin-resolved observables acquire a distinct character. The interplay between non-reciprocal hopping and magnetic exchange therefore raises basic questions regarding the existence, structure, and physical interpretation of persistent spin currents in non-Hermitian systems.

Motivated by these open issues, we investigate persistent spin currents in a non-Hermitian dimerized ring with ferromagnetic exchange, employing the same class of asymmetric hopping previously used to study persistent charge currents~\cite{ganguly2025}. The dimerized Hatano–Nelson ring provides a minimal and controlled platform in which non-reciprocity, topology, and magnetism can be tuned independently, allowing us to isolate the impact of spin degrees of freedom on persistent transport without introducing additional model-specific complexities. By systematically analyzing spin-resolved currents across different parameter regimes, we elucidate how non-Hermiticity and magnetism jointly shape persistent spin transport, thereby extending the understanding of equilibrium-like responses in non-Hermitian quantum systems beyond the spinless case.

Specifically, we study a dimerized HN ring with anti-Hermitian intradimer hopping~\cite{ganguly2025}, incorporating a uniform ferromagnetic ordering to explore spin-resolved persistent currents and their interplay with non-Hermiticity. The ferromagnetic configuration is chosen as the simplest magnetic background that breaks spin degeneracy while preserving translational symmetry, thereby allowing us to isolate the effects of non-Hermiticity on spin-dependent transport without additional complications arising from magnetic frustration or spatially varying textures. The system is modeled within the nearest-neighbor tight-binding framework. Persistent charge and spin currents in mesoscopic rings can be computed through different approaches, including the derivative method~\cite{byers1961,pc2}, the current operator formalism~\cite{spc08,maiti16}, and Green's function techniques~\cite{maiti14}. While the derivative method depends only on the eigenvalues of the Hamiltonian, both the current operator and Green's function approaches require explicit knowledge of the eigenstates. In this work, we adopt the current operator formalism.

For Hermitian systems, it is sufficient to use right eigenstates in the current operator method (or Green's function techniques). However, in non-Hermitian systems the biorthogonal basis must be employed~\cite{shen2024}. The right and left eigenstates and eigenvalues of ${\mathcal H}$ satisfy
$
{\mathcal H} \lvert \psi_\alpha^R\rangle = \epsilon_\alpha \lvert \psi_\alpha^R\rangle
$
and
$
\langle \psi_\alpha^L \rvert {\mathcal H} = \epsilon_\alpha \langle \psi_\alpha^L \rvert,
$
where $\epsilon_\alpha$ is the $\alpha$th eigenvalue, and $\lvert \psi_\alpha^R \rangle$ and $\langle \psi_\alpha^L \rvert$ denote the corresponding right and left eigenstates. For non-Hermitian ${\mathcal H}$, $\langle \psi_\alpha^L \rvert \neq \lvert \psi_\alpha^R \rangle$, making the biorthogonal basis indispensable for obtaining meaningful expectation values of dynamical quantities~\cite{biortho}. With this framework, we develop a detailed theoretical formulation to compute both the  persistent charge current and the $x$, $y$, and $z$ components of the  persistent spin current.

Before proceeding, we briefly clarify the physical meaning of the real and imaginary components of energies and currents that arise in the non-Hermitian framework. In non-Hermitian lattice systems, physical observables evaluated within the biorthogonal framework generally acquire complex values, reflecting the intrinsically non-unitary nature of the dynamics. Accordingly, energy eigenvalues take the form $E_n=E_n^R+iE_n^I$, where the real part $E_n^R$ determines the spectral position of the state and governs equilibrium-like response properties such as flux sensitivity and state occupation, while the imaginary part $E_n^I$ encodes the exponential growth or decay rate of the corresponding mode arising from non-reciprocity, asymmetric hopping, or effective gain-loss processes~\cite{hatano1,hatano2,kawabata}. Persistent charge and spin currents, defined either via the continuity equation or through derivatives of the many-body energy with respect to flux, may likewise possess complex expectation values in non-Hermitian systems. The real part of the persistent charge (spin) current corresponds to the net circulating charge (spin) flow and represents the natural generalization of the Hermitian persistent current, which can in principle be accessed experimentally through magnetic or transport measurements. In contrast, the imaginary part of the current reflects non-conservative probability or spin flow associated with the anti-Hermitian nature of the lattice current operator and has no direct Hermitian analogue. While these imaginary components do not correspond to directly measurable steady-state observables, they carry essential dynamical information and can be indirectly inferred from time-dependent evolution, decay or amplification rates, linewidths, or imbalance of probability flow~\cite{shen2018,longhi2019}. Thus, the appearance of imaginary parts in energies and currents is a genuine physical feature of non-Hermitian transport rather than an artifact of the biorthogonal formalism.

The remainder of the paper is structured as follows. In Section~\ref{sec2}, we introduce the schematic of the ferromagnetic HN ring, the model Hamiltonian, energy dispersion relations, and the theoretical framework required to calculate the charge and all three components of the  persistent spin currents. Section~\ref{sec3} presents the complex energy spectrum, and both real and imaginary parts of the persistent currents under various conditions. Finally, we conclude our main findings in Section~\ref{conclusion}. The detailed mathematical calculations for the Bolch Hamiltonian and dispersion relations, biorthogonalization of eigenstates, and persistent charge and spin currents are provided in different appendices.

\section{\label{sec2}Physical system, Hamiltonian and Theoretical Framework}
\subsection{Schematic and model Hamiltonian}
\begin{figure}[ht]
\centering 
\includegraphics[width=0.45\textwidth]{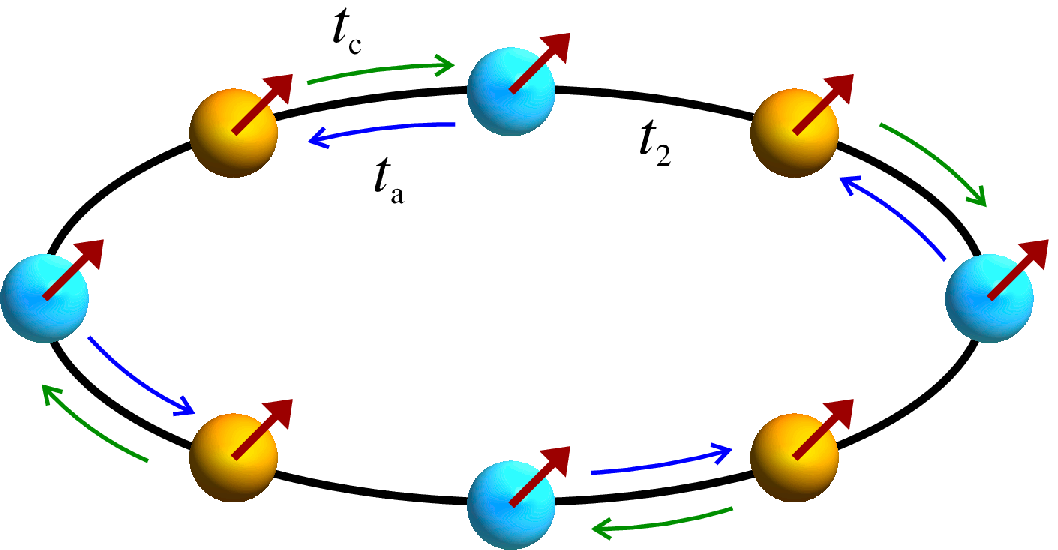}
\caption{(Color online.) Schematic representation of the magnetic Hatano-Nelson ring. The yellow and cyan colored balls correspond to the $A$ and $B$ sublattices, respectively. The red arrows on each site indicate the orientation of the local magnetic moments, which are tilted with respect to the $z$-quantization axis. The intradimer hopping amplitudes are denoted by $t_c$ (clockwise, green) and $t_a$ (counterclockwise, blue), while the interdimer hopping amplitude is represented by $t_2$.}
\label{setup}
\end{figure}

The schematic of an HN ring is illustrated in Fig.~\ref{setup}. Each unit cell of the ferromagnetic ring consists of two magnetic sites, where the yellow and cyan colored balls represent the $A$ and $B$ sublattices, respectively. 
Red arrows on the sites depict the orientation of the local magnetic moments, which are tilted relative to the $z$-quantization axis. The asymmetry is incorporated through the intradimer bond. The corresponding nearest-neighbor tight-binding Hamiltonian under periodic boundary condition (PBC) takes the form~\cite{ganguly2025,moumita-prb,sarkar2019}
\begin{eqnarray}
H &=& \sum_{n=1}^N\sum_{\alpha=A,B}  \bm{c}_{n,\alpha}^\dagger \left(\bm{\epsilon}_{n,\alpha} - \contour[1]{black}{$\mathpzc{h}$}_n \cdot \bm{\sigma} \right) \bm{c}_{n,\alpha} \nonumber \\&+&
\sum\limits_{n=1}^N\left(\bm{c}_{n,A}^{\dagger} \bm{t}_{c} \bm{c}_{n,B} +  \bm{c}_{n,B}^{\dagger} \bm{t}_{a} \bm{c}_{n,A}\right)  \nonumber\\&+&\sum\limits_{n=1}^{N-1}\left(\bm{c}_{n,B}^{\dagger}  \bm{t}_{2} \bm{c}_{n+1,A} + H.c\right) 
\nonumber\\&+&
\left(\bm{c}^\dagger_{N,B} \bm{t}_2 \bm{c}_{1,A} + H.c.\right),
\label{ham}
\end{eqnarray}
where
\begin{align}
&\bm{c}_{n,\alpha} = \begin{pmatrix}c_{n,\alpha \uparrow} \\c_{n,\alpha \downarrow}\end{pmatrix},\quad
\bm{c}_{n,\alpha}^\dagger = \begin{pmatrix}c_{n,\alpha \uparrow}^\dagger & c_{n,\alpha \downarrow}^\dagger\end{pmatrix},\nonumber\\
&\bm{\epsilon}_{n,\alpha} = \begin{pmatrix}\epsilon_{n,\alpha} & 0 \\0 & \epsilon_{n,\alpha}\end{pmatrix},\quad \bm{t}_{c(a)} = \begin{pmatrix} t_{c(a)} & 0\\ 0 & t_{c(a)}\end{pmatrix}, \nonumber\\
&\bm{t}_2 = \begin{pmatrix} t_2 & 0\\ 0 & t_2\end{pmatrix},\quad\contour[1]{black}{$\mathpzc{h}$}_n \cdot \bm{\sigma}  = \mathpzc{h}_n \begin{pmatrix} \cos{\theta_n} & \sin{\theta_n}e^{-i\varphi_n}\\ \sin{\theta_n}e^{i\varphi_n} & -\cos{\theta_n}\end{pmatrix}.\nonumber
\end{align}
Here in Eq.~\ref{ham}, the sum over $n$ runs over unit cells, the sum over $\alpha$ runs over sublattices, and $N$ denotes the total number of unit cells. The operators $c_{n,\alpha \uparrow}$ and $c_{n,\alpha \downarrow}$ represent fermionic annihilation operators on the $\alpha(=A,B)$ sublattice of the $n$th unit cell for the up- and down-spin components, respectively, while $c_{n,\alpha \uparrow}^\dagger$ and $c_{n,\alpha \downarrow}^\dagger$ are the corresponding creation operators.

The first term represents the effective on-site energy with electron-moment interaction. The on-site energy at the $\alpha(=A,B)$ sublattice of the $n$th unit cell is $\epsilon_{n,\alpha}$. The coupling term $\contour[1]{black}{{$\mathpzc{h}$}}_n \cdot \bm{\sigma}$, with $\contour[1]{black}{{$\mathpzc{h}$}}_n = J\langle \mathbf{S}_n \rangle$, defines the spin-dependent scattering (SDS) parameter, where $J$ is the spin-moment exchange coupling and $\langle \mathbf{S}_n \rangle$ is the net spin at the $n$th site~\cite{yhsu}. Here, $\bm{\sigma}={\sigma_x,\sigma_y,\sigma_z}$ are Pauli matrices, while $\theta_n$ and $\varphi_n$ denote the polar and azimuthal angles. With all magnetic moments aligned along a fixed direction relative to the $z$ quantization axis and the SDS parameter taken isotropic, one has $\mathpzc{h}_n=\mathpzc{h}$, $\theta_n=\theta$, and $\varphi_n=\varphi$ for all sites.

The second term describes intradimer hopping, with $t_c$ and $t_a$ denoting the clockwise and anticlockwise amplitudes, respectively. They are non-reciprocal, satisfying $t_a = -t_c^*$, making the Hamiltonian non-Hermitian. Parameterizing $t_c = t + i\gamma$ and $t_a = -t + i\gamma$ ($t,\gamma \in \mathbb{R}$), one obtains the polar form $t_c = \lvert t_1\rvert e^{i\phi}$ and $t_a = -\lvert t_1\rvert e^{-i\phi}$, where $\lvert t_1\rvert=\sqrt{t^2+\gamma^2}$ and $e^{i\phi}$ is the Peierls phase factor. Over a full cycle the accumulated phase is $e^{iN\phi}$, with $\Phi=N\phi$ identified as the effective magnetic flux induced by non-Hermiticity~\cite{li2021,ganguly2025}.

The third term describes the interdimer hopping, which connects the $B$-sublattice of the $n$th unit cell to the $A$-sublattice of the $(n+1)$th unit cell with amplitude $t_2$. This hopping is reciprocal and therefore Hermitian.

The last term connects the first and $N$th unit cells through the hopping amplitude $t_2$, thereby enforcing PBC.

\subsection{Persistent charge current}
The charge current operator is defined as
\begin{equation}
\hat{J} = \frac{e\dot{X}}{2Na}, \quad 
\dot{X} = \frac{2\pi}{ih}[X,H],
\end{equation}
where $e$ is the electronic charge, $a$ is the lattice spacing, and $2N$ is the total number of sites. $X$ and $\dot{X}$ are the position and velocity operators, respectively.  
In the non-Hermitian case, the current for the $n$th eigenstate is evaluated in the biorthogonal basis (a discussion of biorthogonal basis is given in Appendix~\ref{app:biorthogal}) through the operation as  
\begin{equation}
I_n = \left\langle \psi_n^{L} \middle| \hat{J}\middle| \psi_n^{R} \right\rangle,
\end{equation}
where $\lvert \psi_n^R \rangle$ and $\lvert \psi_n^L \rangle$ denote the right and left eigenvectors, respectively.  
The explicit derivation of $\hat{J}$ and its evaluation in terms of the expansion coefficients of $\lvert \psi_n^{R/L} \rangle$ is provided in Appendix~\ref{app:current}.

\subsection{Persistent spin current}
The spin current operator is defined as
\begin{equation}
\hat{I}_\alpha = \frac{\hbar}{4aN} \left( \sigma_\alpha \dot{X} + \dot{X}\,\sigma_\alpha \right),
\quad \alpha \in \{x,y,z\},
\label{def-spin}
\end{equation}
where $\sigma_\alpha$ are the Pauli matrices.  
The spin current in the $n$th eigenstate is evaluated in the biorthogonal basis as
\begin{equation}
I_{\alpha}^n = \left\langle \psi_n^{L} \middle| I_{\mathrm{s}}^\alpha \middle| \psi_n^{R} \right\rangle,
\end{equation}
with $\lvert \psi_n^{R}\rangle$ and $\lvert \psi_n^{L}\rangle$ the right and left eigenstates, respectively.  

The explicit operator forms of $I_x$, $I_y$, and $I_z$, together with their evaluation in terms of the expansion coefficients of $\lvert \psi_n^{R/L} \rangle$, are provided in Appendix~\ref{app:spin_current}.

\section{\label{sec3}Results and Discussion}
Throughout this discussion, all energies are expressed in units of eV, with the interdimer hopping fixed at $t_2 = 1\,$eV. Unless specified, the on-site energy is set as $\epsilon_{n,\uparrow}=\epsilon_{n,\downarrow}=0$ for all the sites. The persistent charge current is reported in units of $\mu$A, whereas the persistent spin current is given in units of eV. Physically, the persistent spin current corresponds to the flow of spin angular momentum per unit time around the ring. Since angular momentum has units of $\hbar$, dividing by time naturally yields units of energy, which justifies expressing the spin current in eV. This is consistent both with the dimensional analysis of Eq.~\ref{def-spin} and with standard tight-binding conventions~\cite{spc07,spc08}. Since the currents are calculated using the current operator method, which relies on the eigenstates of the system Hamiltonian, it is essential to work in the biorthogonal basis, as discussed earlier. To this end, we employ a Gramm-Schmidt-type block diagonalization procedure, which allows us to systematically handle the eigenstates and avoid complications arising from degenerate energy levels.

\subsection{Dispersion relations}
We begin the discussion of our results by analyzing the $k$-space spectrum in detail.
By writing the Hamiltonian in $k$-space in the basis 
\[
[A\uparrow, A\downarrow, B\uparrow, B\downarrow],
\] 
and assuming isotropic on-site potentials $\epsilon_{n,\alpha} = \epsilon$ and spin-dependent scattering $\mathpzc{h}_n = \mathpzc{h}$ with all magnetic moments aligned along a fixed direction ($\theta_n = \theta$, $\varphi_n = \varphi$), the energy eigenvalues of $H(k)$ are
\begin{equation}
E_{s,\alpha}(k+\phi) = \epsilon + s\,\mathpzc{h} + \alpha \sqrt{t_2^2 - \lvert t_1 \rvert^2 + 2 i \lvert t_1 \rvert t_2 \sin(k+\phi)},
\label{dispers}
\end{equation}
where $s = +1$ ($\uparrow_{\mathbf m}$) or $-1$ ($\downarrow_{\mathbf m}$) labels the spin sector along the magnetization axis $\mathbf m$, and $\alpha = -1$ (bonding) or $+1$ (antibonding) denotes the $A-B$ sublattice branch. Accordingly, we label the four bands as 
\[
E_\uparrow^-, \; E_\uparrow^+, \; E_\downarrow^-, \; E_\downarrow^+,
\] 
where the superscript indicates bonding ($-$) or antibonding ($+$) and the subscript indicates spin orientation along $\mathbf m$. The full derivation, starting from the Bloch Hamiltonian and proceeding up to the energy spectrum, is provided in Appendix~\ref{app:kspace}.

Several notable characteristics emerge from the dispersion relations (Eq.~\ref{dispers}). To begin with, the phase $\phi$ and the wave vector $k$ enter the spectrum on an equal footing. Under periodic boundary conditions, the accumulated phase around the ring corresponds to a total factor $e^{iN\phi}$, with $N$ denoting the number of unit cells. Consequently, the non-Hermitian system effectively mimics a real magnetic flux $\Phi = N\phi$ in the ring geometry. Since shifting $\phi$ by $2j\pi/N$ leaves the eigenvalues unchanged, the flux $\Phi$ induces a periodic modulation of the spectrum in the complex energy plane. This gives rise to magneto-oscillations of the energy spectrum, characterized by a period $\Phi = 2\pi$, consistent with earlier studies~\cite{li2021,ganguly2025}. The ferromagnetic ordering introduces a spin splitting of magnitude $\mathpzc{h}$ in the real part of the energy, whereas the imaginary part, originating from the non-Hermitian term $\sqrt{t_2^2 - |t_1|^2 + 2 i |t_1| t_2 \sin(k+\phi)}$, remains identical for both spin channels. Consequently, even for a tilted alignment of the moments ($\theta \neq 0, \varphi \neq 0$), the spin splitting manifests only in the real spectrum, while the imaginary spectrum remains degenerate in spin space.

\begin{figure}[ht]
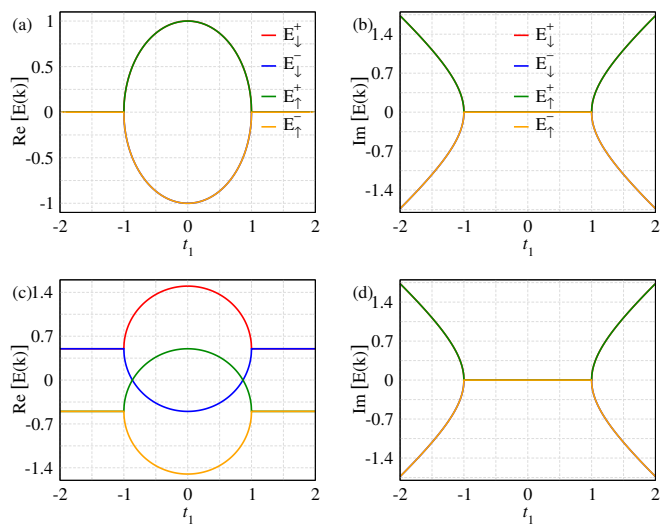
 
\includegraphics[width=0.23\textwidth]{fig2a.eps}\hfill
\includegraphics[width=0.23\textwidth]{fig2b.eps}\vskip 0.1 in
\includegraphics[width=0.23\textwidth]{fig2c.eps}\hfill
\includegraphics[width=0.23\textwidth]{fig2d.eps}
\caption{(Color online.) Complex band structure as a function of $t_1$. Bands are computed using Eq.~\ref{dispers}. (a) Real and (b) imaginary parts of 
$E(k)$ with $\mathpzc{h}=0$. (c) Real and (d) imaginary parts of $E(k)$ with $\mathpzc{h}=0.5$. The colors represent different spin states: red for $E_\downarrow^+$, blue for $E_\downarrow^-$ green for $E_\uparrow^+$, and orange for $E_\uparrow^-$.
}
\label{t1var}
\end{figure}

The $k$-space complex energy spectrum is shown in Fig.~\ref{t1var}, with the upper panel for $\mathpzc{h}=0$ and the lower panel for $\mathpzc{h}=0.5$. The on-site potential 
is set to $\epsilon_{n,\alpha}=0$, and the interdimer hopping is fixed at $t_2=1$. For $\mathpzc{h}=0$, the system behaves like a conventional non-Hermitian SSH model, where both real and imaginary spectra are spin-degenerate. Accordingly, the red ($E_\downarrow^+$) and green ($E_\uparrow^+$) curves overlap at higher energies (antibonding), while the blue ($E_\downarrow^-$) and orange ($E_\uparrow^-$) curves overlap at lower energies (bonding).

In the real spectrum (Fig.~\ref{t1var}), a gap exists for $\lvert t_1\rvert \leq t_2$, representing the topological phase~\cite{li2021,ganguly2025}. At $\lvert t_1\rvert=t_2$, the gap closes at the critical point, and for $\lvert t_1\rvert>1$, the spectrum is gapless, indicating the trivial phase. The imaginary spectrum shows the opposite trend. It is gapless in the topological phase, gap occurs at the critical point, and remain gapped in the trivial phase (Fig.~\ref{t1var}(b)).

When ferromagnetic ordering is included, the real spectrum splits into spin-up and spin-down components (Fig.~\ref{t1var}(c)). The bonding-antibonding gap is preserved (red and blue for spin-down, green and orange for spin-up), but spin splitting modifies the energy separation between spins. The imaginary spectrum  (Fig.~\ref{t1var}(d)) remains unchanged, identical to the spinless case, since the ferromagnetic ordering does not affect the imaginary space as discussed earlier.

\begin{figure}[ht]
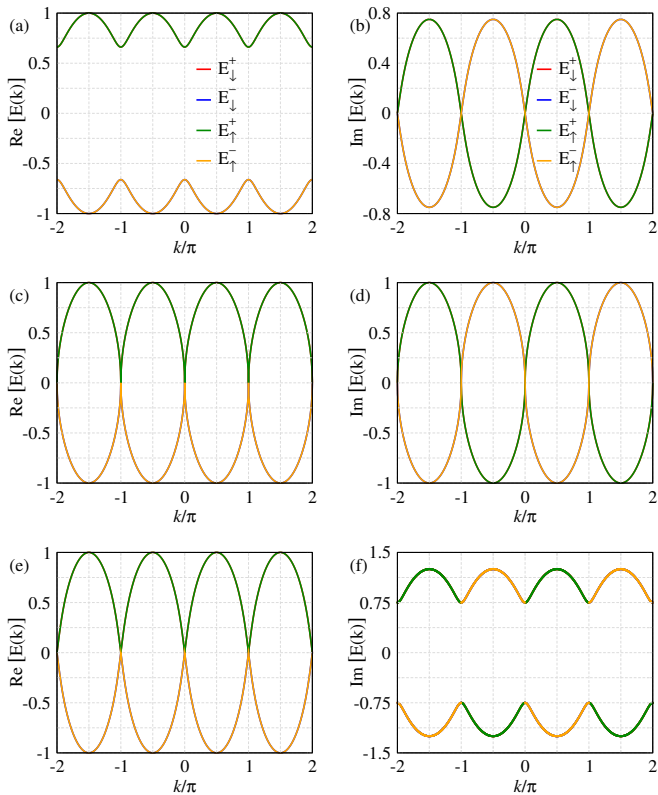
 
\includegraphics[width=0.23\textwidth]{fig3a.eps}\hfill
\includegraphics[width=0.23\textwidth]{fig3b.eps}\vskip 0.1 in
\includegraphics[width=0.23\textwidth]{fig3c.eps}\hfill
\includegraphics[width=0.23\textwidth]{fig3d.eps}\vskip 0.1 in
\includegraphics[width=0.23\textwidth]{fig3e.eps}\hfill
\includegraphics[width=0.23\textwidth]{fig3f.eps}
\caption{(Color online.) Complex energy spectra $E(k)$ in momentum space for three representative values of the intradimer hopping amplitude $t_1$, with fixed interdimer hopping $t_2 = 1$ and spin-dependent scattering parameter $\mathpzc{h} = 0$. Panels (a-b), (c-d), and (e-f) correspond to $\lvert t_1 \rvert = 0.75$, $1.0$, and $1.25$, respectively. Left column (a, c, e): real part of the spectrum $\mathrm{Re}[E(k)]$; right column (b, d, f): imaginary part $\mathrm{Im}[E(k)]$. The color codes for the bands are identical to those in Fig.~\ref{t1var}.}
\label{kkk}
\end{figure}
\begin{figure}[ht]
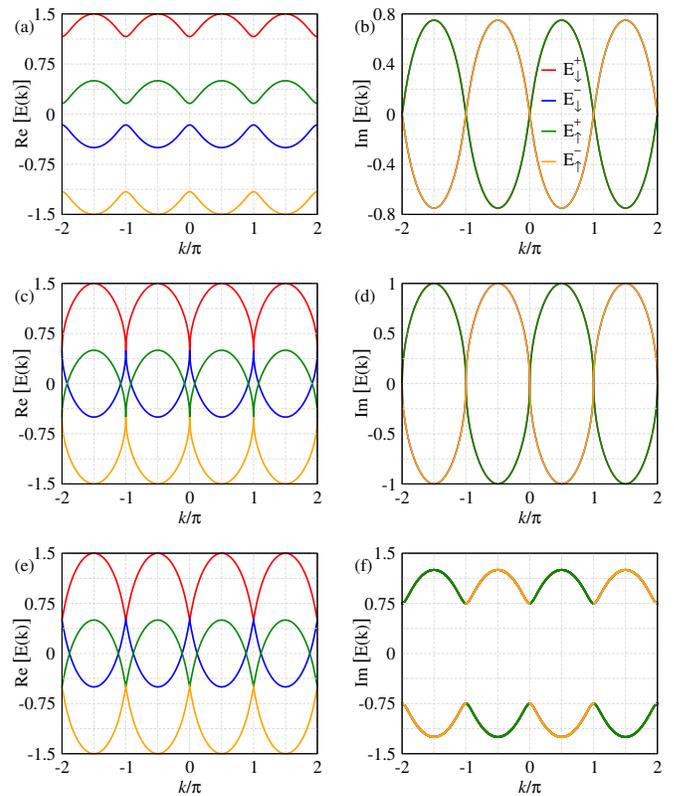
 
\includegraphics[width=0.23\textwidth]{fig4a.eps}\hfill
\includegraphics[width=0.23\textwidth]{fig4b.eps}\vskip 0.1 in
\includegraphics[width=0.23\textwidth]{fig4c.eps}\hfill
\includegraphics[width=0.23\textwidth]{fig4d.eps}\vskip 0.1 in
\includegraphics[width=0.23\textwidth]{fig4e.eps}\hfill
\includegraphics[width=0.23\textwidth]{fig4f.eps}
\caption{(Color online.) Complex energy spectra $E(k)$ in momentum space for three representative values of the intradimer hopping amplitude $t_1$, with fixed interdimer hopping $t_2 = 1$ and spin-dependent scattering parameter $\mathpzc{h} = 0.5$. Panels (a-b), (c-d), and (e-f) correspond to $\lvert t_1 \rvert = 0.75$, $1.0$, and $1.25$, respectively. Left column (a, c, e): real part of the spectrum $\mathrm{Re}[E(k)]$; right column (b, d, f): imaginary part $\mathrm{Im}[E(k)]$. The color codes for the bands are identical to those in Fig.~\ref{t1var}.}
\label{kkkh}
\end{figure}

To elucidate the topological and spectral properties of the model, we examine the complex band structure in momentum space for three representative values of the intradimer hopping amplitude $t_1$, while keeping the interdimer hopping fixed at $t_2 = 1$. The results are shown in Fig.~\ref{kkk} for $\mathpzc{h} = 0$. All other parameters are identical to those in Fig.~\ref{t1var}, and the color coding of the bands follows the same convention.

In the upper panel (Figs.~\ref{kkk}(a) and (b)), corresponding to $\lvert t_1 \rvert = 0.75$, the system lies in the topological phase, analogous to the non-Hermitian Hatano-Nelson model~\cite{li2021,ganguly2025}. The real 
part of the spectrum (Fig.~\ref{kkk}(a)) exhibits a clear bulk energy gap, characteristic of a topologically nontrivial band structure. The imaginary part (Fig.~\ref{kkk}(b)), however, is gapless, the bands are smoothly connected across the Brillouin zone without any separation in their imaginary components.

In the middle panel (Figs.~\ref{kkk}(c) and (d)), the system is at the critical point $\lvert t_1 \rvert = 1$, where the topological phase transition occurs. The real part of the spectrum (Fig.~\ref{kkk}(c)) becomes gapless, signaling the closing of the topological real energy gap. This marks the transition between the topological and trivial phases. The imaginary part (Fig.~\ref{kkk}(d)) is continuous and exhibits no abrupt spectral features. Both the real and imaginary parts of the spectrum are identical to one another, consistent with the dispersion relation given in Eq.~\ref{dispers}.

In the lower panel (Figs.~\ref{kkk}(e) and (f)), with $\lvert t_1 \rvert = 1.25$, the system enters the trivial phase, as defined by the SSH classification. However, unlike in the Hermitian SSH model, the real part of the spectrum (Fig.~\ref{kkk}(e)) is gapless. The upper and lower bands touch at high-symmetry points $k = 0, \pm\pi$. This indicates that the presence of anti-Hermitian intradimer hopping modifies the trivial phase such that it no longer supports a real energy gap. The imaginary part of the spectrum (Fig.~\ref{kkk}(f)) exhibits well-separated bands at $k = 0, \pm\pi, \pm 2\pi$, and so on, reflecting stronger non-Hermitian effects.

All the spectra exhibit periodicity in momentum $k$ with a period of $2\pi$, consistent with the dispersion relation given in Eq.~\ref{dispers}. Throughout all panels, since the SDS parameter is set to $\mathpzc{h} = 0$, no spin splitting is observed. As a result, $E^\pm_\uparrow = E^\pm_\downarrow$, and the corresponding bands exactly overlap. The role of $\mathpzc{h}$ in lifting the spin degeneracy and inducing additional spectral structure will be explored in the following section.

\begin{figure*}[ht]
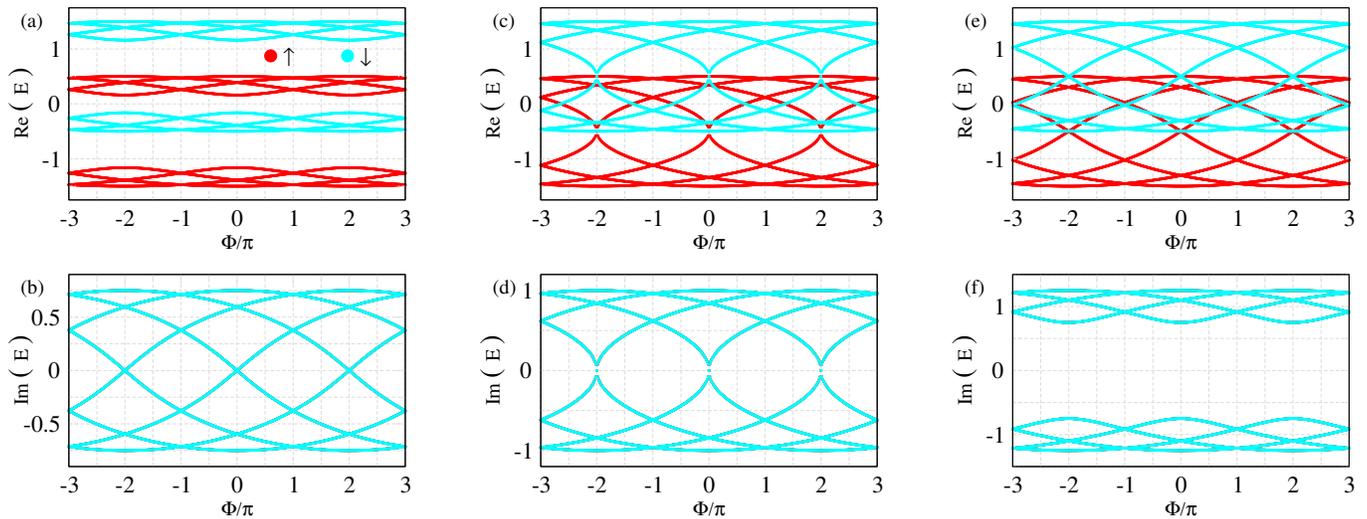
 
\includegraphics[width=0.3\textwidth]{fig5a.eps}\hfill
\includegraphics[width=0.3\textwidth]{fig5c.eps}\hfill
\includegraphics[width=0.3\textwidth]{fig5e.eps}\vskip 0.1 in
\includegraphics[width=0.3\textwidth]{fig5b.eps}\hfill
\includegraphics[width=0.3\textwidth]{fig5d.eps}\hfill
\includegraphics[width=0.3\textwidth]{fig5f.eps}
\caption{(Color online.) Real and imaginary energy eigenspectra as a function of the normalized magnetic flux $\Phi/\pi$. The first, second, and third columns correspond to $t_1 = 0.75$, $1$, and $1.25$, respectively. The number of unit cells is $N = 8$ and the interdimer hopping strength is $t_2 = 1$, with $\mathpzc{h} = 0.5$. Red and cyan colors denote the up- and down-spin sectors, respectively.}
\label{ephi}
\end{figure*}
\begin{figure*}[ht]
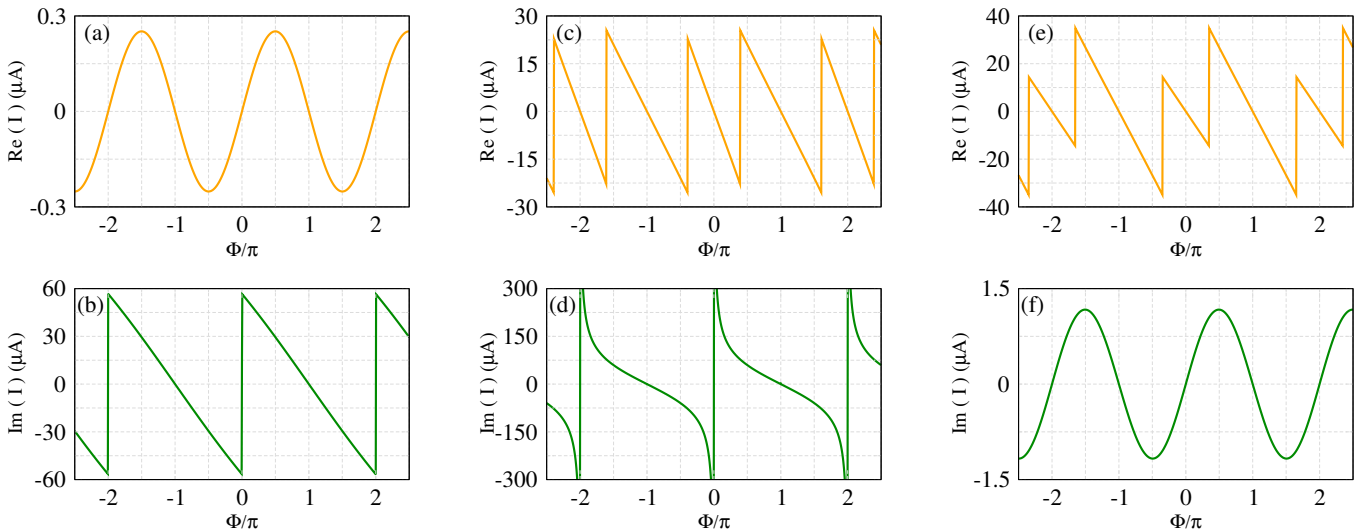
 
\includegraphics[width=0.3\textwidth]{fig6a.eps}\hfill
\includegraphics[width=0.3\textwidth]{fig6c.eps}\hfill
\includegraphics[width=0.3\textwidth]{fig6e.eps}\vskip 0.1 in
\includegraphics[width=0.3\textwidth]{fig6b.eps}\hfill
\includegraphics[width=0.3\textwidth]{fig6d.eps}\hfill
\includegraphics[width=0.3\textwidth]{fig6f.eps}
\caption{(Color online.) Real and imaginary persistent charge current as a function of the normalized magnetic flux $\Phi/\pi$. The first, second, and third columns correspond to $t_1=0.75$, $1$, and $1.25$, respectively. The number of unit cells is $N=20$, the interdimer hopping amplitude is $t_2=1$, the spin-dependent splitting is $\mathpzc{h}=0.5$, and the chemical potential is $\mu=0$.}
\label{iphi}
\end{figure*}
The dispersion relation in Fig.~\ref{kkkh} illustrates the combined effects of spin splitting and non-Hermiticity in the ferromagnetic Hatano-Nelson ring. The exchange field $\mathpzc{h}(=0.5)$ induces 
a rigid spin splitting in the real spectra: the up-spin bands (green and orange) are shifted downward relative to the down-spin bands (red and blue) by $2\mathpzc{h}$, while the imaginary parts remain unchanged as $\mathpzc{h}=0$ case since they are insensitive to spin-splitting effect. Within each spin channel, the bonding ($-$) and antibonding ($+$) branches are separated by the square-root term of Eq.~\ref{dispers}, and the behavior of this separation depends on the hopping ratio $|t_1|/t_2$. In the topological regime $|t_1|<t_2$ (Fig.~\ref{kkkh}(a)), a finite real gap persists between bonding and antibonding states for both spin species, yielding four well-separated dispersions that are offset vertically by spin and horizontally by sublattice hybridization. At the critical point $|t_1|=t_2$ (Fig.~\ref{kkkh}(c)), this real gap closes at the high-symmetry points $k=0,\pm\pi,\pm2\pi$. In the trivial regime $|t_1|>t_2$ (Fig.~\ref{kkkh}(e)), the real-part gap collapses and the bonding-antibonding distinction is transferred to the imaginary channel. Overall, the exchange field $\mathpzc{h}$ simply shifts the two spin manifolds relative to each other without altering the non-Hermitian structure of the bonding-antibonding separation, while the ratio $|t_1|/t_2$ dictates whether the gap appears in the real or imaginary parts of the spectrum. Since the imaginary spectra are identical to those in the absence of the SDS parameter case, they are not discussed here but the plots are included for completeness.

\subsection{Real-space analysis: Persistent charge and spin currents}
To develop a deeper physical understanding of these results, we next examine a simplified real-space configuration. Specifically, we consider the case where all magnetic moments are aligned along the $z$-quantization axis, i.e., $\theta = 0$ and $\phi = 0$. In this limit, the Hamiltonian (Eq.~\ref{ham}) decouples into independent up- and down-spin sectors, $\left(H_\uparrow + H_\downarrow\right)$. We set the SDS parameter to $\mathpzc{h}=0.5$ and fix the number of unit cells in the Hatano-Nelson ring at $N=8$. The resulting complex energy spectra for the up- and down-spin Hamiltonians are shown as a function of $\Phi$ in Fig.~\ref{ephi}, for three representative values of the intradimer hopping amplitude. Since the hopping amplitudes follow an SSH-type dimerization, we examine three characteristic regimes: the topological phase $\lvert t_1\rvert < t_2$, the critical point $\lvert t_1\rvert = t_2$, and the trivial phase $\lvert t_1\rvert > t_2$.

\begin{figure*}[ht]
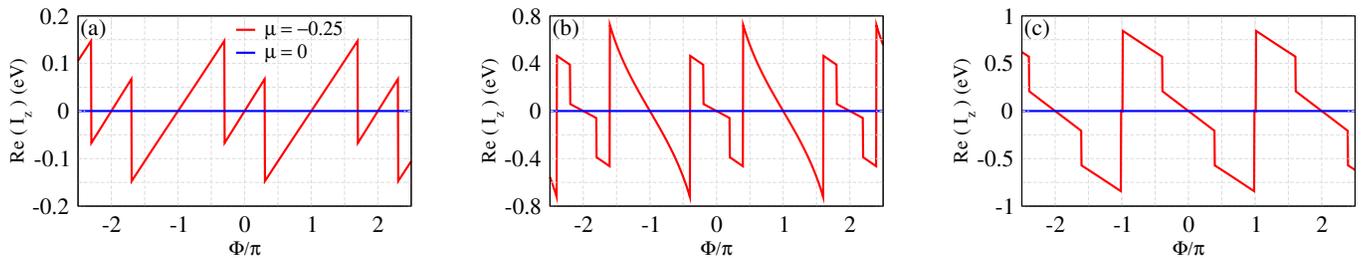
 
\includegraphics[width=0.3\textwidth]{fig7a.eps}\hfill
\includegraphics[width=0.3\textwidth]{fig7b.eps}\hfill
\includegraphics[width=0.3\textwidth]{fig7c.eps}
\caption{(Color online.) Real part of the z-component of the  persistent spin current, Re($I^z_s$), as a function of the normalized magnetic flux $\Phi/\pi$ for (a) $\lvert t_1\rvert = 0.75$, (b) $\lvert t_1\rvert = 1$, and (c) $\lvert t_1\rvert = 1.25$. The system parameters are fixed at number of unit cells $N = 20$, interdimer hopping strength $t_2 = 1$, and SDS paramer $\mathpzc{h} = 0.5$. Results are shown for two chemical potential values, $\mu=0$ and -0.25 with blue and red colors, respectively.}
\label{izphi}
\end{figure*}
\begin{figure*}[ht]
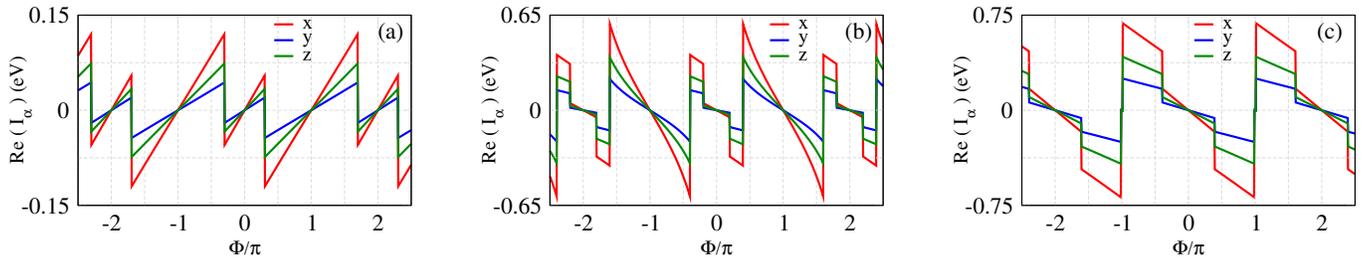
 
\includegraphics[width=0.3\textwidth]{fig8a.eps}\hfill
\includegraphics[width=0.3\textwidth]{fig8b.eps}\hfill
\includegraphics[width=0.3\textwidth]{fig8c.eps}
\caption{(Color online.) Real $x$, $y$, and $z$ components of the persistent spin current as a function of the normalized magnetic flux $\Phi/\pi$ for (a) $\lvert t_1\rvert = 0.75$, (b) $\lvert t_1\rvert = 1$, and (c) $\lvert t_1\rvert = 1.25$. The system parameters are $N = 20$, $t_2 = 1$, $\mathpzc{h} = 0.5$, $\theta = \pi/3$, $\varphi = \pi/9$, and $\mu = -0.25$. Red, blue, and green curves represent the results for the  $x$, $y$, and $z$ components, respectively.
}
\label{xyzphi}
\end{figure*}

The complex energy spectrum shows that the SDS parameter $\mathpzc{h}$ separates the up- and down-spin energies in the real part, while the imaginary part remains spin-degenerate. All the spectra exhibit an oscillation with $\Phi$ and a period of $2\pi$. These behaviors reflect the dispersion relation where ferromagnetic ordering produces a real-energy splitting of magnitude $\mathpzc{h}$, while the imaginary component arising from the anti-Hermitian intradimer hopping is identical for both spin channels. Apart from the shift in the energy window, all characteristic features of the up- and down-spin real spectra for different values of $\lvert t_1\rvert$ remain the same as in the spinless case. The modification of the energy window in the real spectra results from the presence of $\mathpzc{h}$. The imaginary spectra are completely identical to the spinless case since ferromagnetic ordering does not affect the imaginary part. Here the imaginary spectra of the up and down spins completely overlap. A detailed discussion of the complex energy spectra is given in Refs.~\cite{li2021,ganguly2025} and here only the key points are mentioned.

In the topological phase ($\lvert t_1\rvert < t_2$), the real spectra for both up and down spins exhibit a gap (Fig.~\ref{ephi}(a)). In the spinless case this gap appears as a line gap around zero energy. With $\mathpzc{h}$ the gap shifts away from zero in both spin channels. The imaginary eigenvalues evolve continuously with $\Phi$ (Fig.~\ref{ephi}(b)) and the spectra remain gapless for both up- and down-spin cases.

At the critical point ($\lvert t_1\rvert = t_2$), the real spectra for both up and down spins (Fig.~\ref{ephi}(c)) are gapless, and the imaginary spectra (Fig.~\ref{ephi}(d)) are also gapless. In the spinless case the real and imaginary spectra coincide, whereas in the FM ring a splitting appears in the real part. Apart from the shift in the energy window, the up- and down-spin real spectra are identical to each other and also coincide with the corresponding up- and down-spin imaginary spectra.

In the trivial phase ($\lvert t_1\rvert > t_2$), the situation is opposite to that in the topological phase. The imaginary spectra (Fig.~\ref{ephi}(f)) are gapped, while the up- and down-spin real spectra (Fig.~\ref{ephi}(f)) remain gapless.

Next, we compute the persistent charge current using Eq.~\ref{charge-curr}, considering the contributions of the energy levels up to the chemical potential $\mu=0$. The number of unit cells is $N=20$. The SDS parameter is fixed at $\mathpzc{h}=0.5$, with $\theta=0$ and $\varphi=0$. Under these conditions the system Hamiltonian decouples into independent up- and down-spin Hamiltonians, as in the energy spectrum analysis.

\begin{figure*}[ht] 
\includegraphics[width=0.3\textwidth]{fig9a.eps}\hfill
\includegraphics[width=0.3\textwidth]{fig9b.eps}\hfill
\includegraphics[width=0.3\textwidth]{fig9c.eps}
\caption{(Color online.) Real $x$, $y$, and $z$ components of the persistent spin current as a function of the normalized polar angle $\theta/\pi$ for (a) $\lvert t_1\rvert = 0.75$, (b) $\lvert t_1\rvert = 1$, and (c) $\lvert t_1\rvert = 1.25$. The flux is fixed at $\Phi = \pi/4$. The system parameters are $N = 20$, $t_2 = 1$, $\mathpzc{h} = 0.5$, $\varphi = \pi/9$, and $\mu = -0.25$. Red, blue, and green curves denote the $x$, $y$, and $z$ components, respectively.
}
\label{thetavar}
\end{figure*}
\begin{figure*}[ht] 
\includegraphics[width=0.3\textwidth]{fig10a.eps}\hfill
\includegraphics[width=0.3\textwidth]{fig10b.eps}\hfill
\includegraphics[width=0.3\textwidth]{fig10c.eps}
\caption{(Color online.) Real $x$, $y$, and $z$ components of the persistent spin current as a function of the normalized azimuthal angle $\varphi/\pi$ for (a) $\lvert t_1\rvert = 0.75$, (b) $\lvert t_1\rvert = 1$, and (c) $\lvert t_1\rvert = 1.25$. The flux is fixed at $\Phi = \pi/4$. The system parameters are $N = 20$, $t_2 = 1$, $\mathpzc{h} = 0.5$, $\theta = \pi/3$, and $\mu = -0.25$. Red, blue, and green curves denote the $x$, $y$, and $z$ components, respectively.}
\label{phivar}
\end{figure*}
Figure~\ref{iphi} shows the real and imaginary parts of the persistent charge current as a function of $\Phi$ for three representative values of the intradimer hopping $t_1$. In the topological phase ($\lvert t_1\rvert =0.75$) the real part (Fig.~\ref{iphi}(a)) exhibits a weak sinusoidal oscillation, while the imaginary part (Fig.~\ref{iphi}(b)) dominates with a sawtooth-like profile and sharp discontinuities at $\Phi=0,\pm 2\pi$. At the critical point ($\lvert t_1\rvert=1$) both the real (Fig.~\ref{iphi}(c)) and imaginary (Fig.~\ref{iphi}(d)) components show enhanced magnitudes, consistent with the gapless spectrum. An important feature here is that the real and imaginary currents are no longer identical, unlike the spinless case, which is also evident from the energy spectrum at the critical point (Figs.~\ref{ephi}(c) and (d)). The imaginary current is discontinuous at $\Phi=0,\pm 2\pi$. In the trivial phase ($\lvert t_1\rvert=1.25$) the real part (Fig.~\ref{iphi}(e)) consists of piecewise-linear segments separated by discontinuous jumps, and the imaginary part (Fig.~\ref{iphi}(f)) is weak and smooth with small-amplitude oscillations. Overall, the persistent charge current oscillates with $\Phi$ with a period of $2\pi$, and the relative strength of the real and imaginary components clearly distinguishes the topological, critical, and trivial regimes.

Before presenting the results for the different components of the persistent spin current, it should be noted that all components, $\left(I_x, I_y, I_z\right)$, are identically zero at $\mu=0$. This behavior arises from two key conditions. First, the orbital Hamiltonian must possess \textit{particle-hole (chiral) symmetry}, ensuring that each negative-energy state has a partner at positive energy with opposite orbital current. Second, the hopping terms must be \textit{spin-independent}, with no spin-flip processes or spin-orbit coupling. In the model considered here, both conditions are satisfied, so the spin-up and spin-down contributions within each occupied negative-energy state cancel exactly, yielding $I_x = I_y = I_z = 0$ at $\mu=0$, regardless of the orientation of the local magnetization. If either condition is broken, for example, by disorder, sublattice potentials, or spin-flip hoppings, the cancellation no longer occurs, and the spin current can become finite even at $\mu=0$.

In the present case, with magnetic moments aligned along the $z$-axis, the eigenstates are spin-diagonal, so the expectation values of $\sigma_x$ and $\sigma_y$ vanish. Consequently, only the $z$-component of the persistent spin current can be non-zero, while $I_x$ and $I_y$ remain zero for all chemical potentials. Therefore, we present here the behavior of the real $z$-component of the persistent spin current as a function of $\Phi$, as shown in Fig.~\ref{izphi}, for three representative values of $\lvert t_1 \rvert$, as before. Here $\mu=0$ is considered for completeness, while a non-zero $I_z$ can be observed for $\mu=-0.25$, and their corresponding results are shown with blue and red, respectively. The imaginary part of $I_z$ is not shown, as there is no spin-splitting effect in the imaginary space, as discussed previously. The current oscillates in a sawtooth-like pattern between approximately $\pm 0.2\,$eV in the topological regime (Fig.~\ref{izphi}(a)). At the critical point, the sawtooth profile becomes much more pronounced, with the oscillation amplitude increasing to about $\pm 0.8\,$eV. The discontinuities in the red curve occur at the same positions as in the topological case. In the trivial phase, the current still exhibits sawtooth oscillations, with the amplitude reaching up to $\pm 1\,$eV. Compared to the previous two cases, the number of discontinuities is reduced. In all regimes, $I_z$ shows periodic oscillations with $\Phi$, with a period of $2\pi$.

Now we tilt the magnetic moments with respect to the $z$-quantization axis by setting the polar angle $\theta=\pi/3$ and the azimuthal angle $\varphi=\pi/9$. Under this condition, the system Hamiltonian can no longer be decoupled into independent up- and down-spin Hamiltonians, resulting in spin mixing. Consequently, one expects nonzero $x$- and $y$-components of the persistent spin current in addition to the $z$-component. The SDS parameter, chemical potential, and all other system parameters are kept unchanged as before. The corresponding results are shown in Fig.~\ref{xyzphi}. The overall spectra of the three components mimic the patterns observed in their respective phases in Fig.~\ref{izphi}, but with different magnitudes.

\begin{figure*}[ht] 
\includegraphics[width=0.3\textwidth]{fig11a.eps}\hfill
\includegraphics[width=0.3\textwidth]{fig11b.eps}\hfill
\includegraphics[width=0.3\textwidth]{fig11c.eps}
\caption{(Color online.) Real $x$, $y$, and $z$ components of the persistent spin current as a function of the chemical potential $\mu$ for (a) $\lvert t_1\rvert = 0.75$, (b) $\lvert t_1\rvert = 1$, and (c) $\lvert t_1\rvert = 1.25$. The flux is fixed at $\Phi = \pi/4$. The system parameters are $N = 20$, $t_2 = 1$, $\mathpzc{h} = 0.5$, $\theta = \pi/3$, $\varphi=\pi/9$,  and $\mu = -0.25$. Red, blue, and green curves denote the $x$, $y$, and $z$ components, respectively. }
\label{muvar}
\end{figure*}
\begin{figure*}[ht]
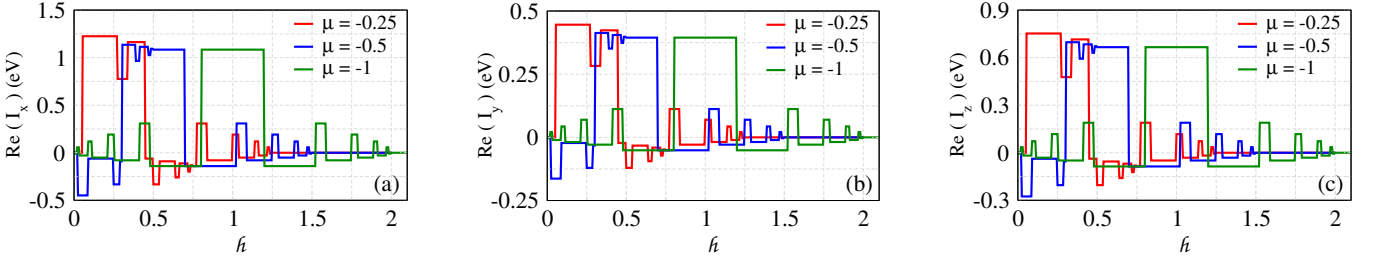
 
\includegraphics[width=0.3\textwidth]{fig12a.eps}\hfill
\includegraphics[width=0.3\textwidth]{fig12b.eps}\hfill
\includegraphics[width=0.3\textwidth]{fig12c.eps}
\caption{(Color online.) Real (a) $x$, (b) $y$, and (c) $z$-components of the persistent spin current as a function of $\mathpzc{h}=0.5$ at the critical point. Here, the number of unit cells $N=20$. The intracell and intercell hopping strengths are identical that is $\lvert t_1\rvert=t_2=1$. $\Phi=0.25$.  $\theta=\pi/3$, $\phi=\pi/9$. Three different chemical potentials are considered, $\mu= -1,-0.5,$ and $-0.25$ and the corresponding results are denoted with green, blue, and red colors, respectively.}
\label{hvar}
\end{figure*}

In the topological phase (Fig.~\ref{xyzphi}(a)), all three components display sawtooth-like oscillations between approximately $\pm 0.15\,$eV. The components are closely spaced, with $I_x$ exhibiting the largest amplitude and $I_z$ the smallest. The discontinuities occur periodically within the $\Phi$-window. At the critical point (Fig.~\ref{xyzphi}(b)), the oscillation amplitude increases significantly, with $I_x$ reaching about $\pm 0.65\,$eV. The sawtooth profile becomes sharper, and the discontinuities occur at the same flux values as in Fig.~\ref{xyzphi}(a). Although separated in magnitude, the three components follow similar trends. In the trivial phase (Fig.~\ref{xyzphi}(c)), the sawtooth oscillations persist, with $I_x$ spanning about $\pm 0.75\,$eV. Compared to the previous two cases, the number of discontinuities per flux period is reduced. The $x$-component continues to dominate, while the $y$- and $z$-components remain smaller and close to one another. Across all cases, the persistent spin currents remain periodic in $\Phi$ with period $2\pi$.

The behavior of the persistent spin currents with the polar angle $\theta$ is examined in Fig.~\ref{thetavar}. Here we use the same set of parameters to compute the three components of the persistent spin current. Certainly, tilting the spin quantization axis modifies the distribution of spin currents among the $x$, $y$, and $z$ components. The first observation is that $I_z$ is antisymmetric about $\theta=\pi$, while $I_x$ and $I_y$ are symmetric. The antisymmetric nature of the $z$-component arises as follows. At $\theta=0$, all magnetic moments are aligned along the positive $z$ direction, leading to a maximum $I_z$, while $I_x$ and $I_y$ vanish, as confirmed by their identically zero values in Fig.~\ref{thetavar}. As $\theta$ increases toward $\pi/2$, the configuration resembles that of spherical polar coordinates, where the moments lie entirely in the $x$-$y$ plane (with nonzero $\varphi$). Consequently, $I_z=0$ but $I_x$ and $I_y$ become finite. At $\theta=\pi$, the moments align along the negative $z$ direction, reversing the sign of $I_z$ while retaining the same magnitude, and again $I_x$ and $I_y$ vanish. This explains the antisymmetric profile of $I_z$. In contrast, the symmetric nature of $I_x$ and $I_y$ arises because, in spherical polar coordinates, they scale as $\sin{\theta}$ for a given $\varphi$, making them symmetric about $\theta=\pi/2$.

Turning to the magnitudes, we observe the following. In the topological phase (Fig.~\ref{thetavar}(a)), all components are small, within $\pm 0.06\,$eV, reflecting the weak current response in this regime. At the critical point (Fig.~\ref{thetavar}(b)), the amplitudes of all three components grow significantly, with $I_z$ spanning nearly $\pm 0.5\,$eV. In the trivial phase (Fig.~\ref{thetavar}(c)), the overall trend resembles the earlier cases but with reduced amplitudes (about $\pm 0.15\,$eV) compared to the critical point.

Another key factor determining the orientation of the magnetic moments is the azimuthal angle $\varphi$, whose influence on the persistent spin currents is illustrated in Fig.~\ref{phivar}. Here the polar angle is fixed at $\theta=\pi/3$, with all other parameters unchanged. For all intradimer hopping configurations, $I_z$ remains constant over the full range of $\varphi$, while $I_x$ shows a cosine-like variation and $I_y$ exhibits a sine-like dependence. This behavior can be physically understood from the spherical polar decomposition of the magnetic moments. For a fixed $\theta$, the $z$-component is independent of $\varphi$, which explains the flat profile of $I_z$, whereas the $x$- and $y$-components scale as $\cos{\varphi}$ and $\sin{\varphi}$, respectively, giving rise to the observed oscillations in $I_x$ and $I_y$.

In terms of magnitudes, the topological phase (Fig.~\ref{phivar}(a)) again shows weak response, in close analogy with the $\theta$-variation (Fig.~\ref{thetavar}(a)). Here $I_z$ is pinned near $0.03\,$eV, while $I_x$ and $I_y$ oscillate within $\pm 0.05\,$eV. At the critical point (Fig.~\ref{phivar}(b)), the magnitudes are strongly enhanced. $I_z$ remains fixed at about $-0.2\,$eV, and $I_x$ and $I_y$ extend almost up to $\pm 0.4\,$eV. In the trivial phase (Fig.~\ref{phivar}(c)), the magnitudes are reduced once again, paralleling the $\theta$-variation (Fig.~\ref{thetavar}(c)). Here $I_z$ stays constant at about $0.06\,$eV, while $I_x$ and $I_y$ oscillate within $\pm 0.12\,$eV. Taken together with the $\phi$-dependence and the $\theta$-variation, these results highlight that the spin persistent current is strongly modulated by the orientation of the magnetic moments and respond anisotropically in spin space.

Figure~\ref{muvar} presents the real parts of the persistent spin current components, $I_x$, $I_y$, and $I_z$, as a function of the chemical potential $\mu$ for three representative intradimer hopping strengths, with all other parameters kept unchanged. Across all cases, the $x$-component is the most dominant, followed by $z$, while the $y$-component remains the weakest. A prominent feature in all phases is the staircase-like profile of the currents. These structures arise because the persistent spin current is directly tied to the occupancy of discrete energy levels. As $\mu$ is varied, each time it crosses an eigenvalue, a new state contributes to the current, producing sharp jumps. The overall dependence on $t_1$ is consistent with the earlier results. In the topological phase (Fig.~\ref{muvar}(a)), the spin current response is weak. At the critical point (Fig.~\ref{muvar}(b)), all three components undergo strong amplification. In the trivial phase (Fig.~\ref{muvar}(c)), the response is intermediate. Here the currents are suppressed relative to the critical point but remain larger than those in the topological regime. Another key feature is that all components across all phases are antisymmetric about $\mu=0$. This antisymmetry originates from particle-hole symmetry and implies that the direction of spin flow reverses when switching from electron-like filling ($\mu>0$) to hole-like filling ($\mu<0$).

\begin{figure*}[ht]
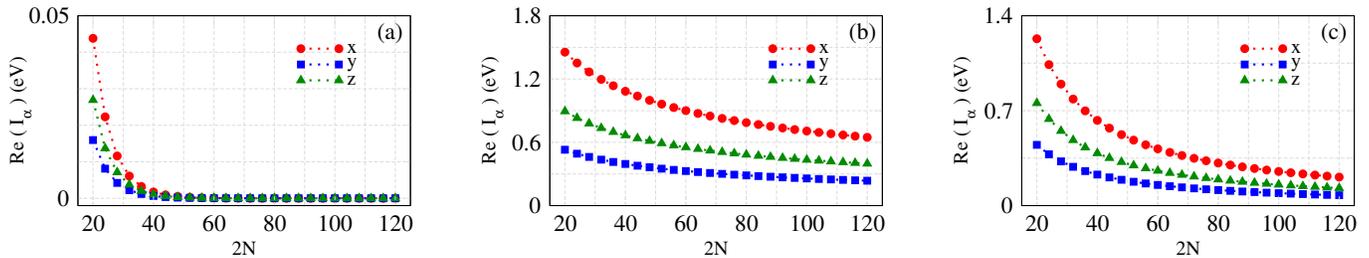
 
\includegraphics[width=0.3\textwidth]{fig13a.eps}\hfill
\includegraphics[width=0.3\textwidth]{fig13b.eps}\hfill
\includegraphics[width=0.3\textwidth]{fig13c.eps}
\caption{(Color online.) Real $x$, $y$, and $z$ components of the spin persistent current as a function of the number of sites $2N$, where $N$ is the number of unit cells. Panels correspond to (a) $\lvert t_1\rvert = 0.75$, (b) $\lvert t_1\rvert = 1$, and (c) $\lvert t_1\rvert = 1.25$. The interdimer hopping strength is $t_2 = 1$, and the magnetic flux is fixed at $\Phi = \pi/4$. The SDS parameter is $\mathpzc{h} = 0.5$, with $\theta = \pi/3$ and $\phi = \pi/9$. The chemical potential is set to $\mu = -0.5$. Red, blue, and green curves correspond to the $x$, $y$, and $z$ components of the persistent spin current, respectively.}
\label{sysvar}
\end{figure*}

\begin{figure*}[ht]
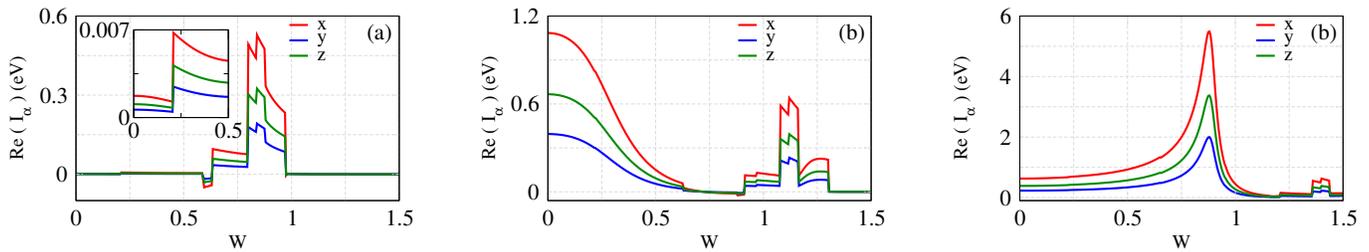
 
\includegraphics[width=0.3\textwidth]{fig14a.eps}\hfill
\includegraphics[width=0.3\textwidth]{fig14b.eps}\hfill
\includegraphics[width=0.3\textwidth]{fig14c.eps}
\caption{Color online.) Real $x$, $y$, and $z$ components of the spin persistent current as a function of the AAH modulation strength $W$. Panels correspond to (a) $\lvert t_1\rvert = 0.75$, (b) $\lvert t_1\rvert = 1$, and (c) $\lvert t_1\rvert = 1.25$. The system has $N = 20$ unit cells, with intercell hopping $t_2 = 1$ and magnetic flux $\Phi = 0.25$. The SDS parameter is $\mathpzc{h} = 0.5$, with $\theta = \pi/3$ and $\phi = \pi/9$. The chemical potential is set to $\mu = -0.5$. Red, blue, and green curves represent the results for the $x$, $y$, and $z$ components of the persistent spin current, respectively.}
\label{disvar}
\end{figure*}

The generation of persistent spin currents with the SDS parameter $\mathpzc{h}$ is investigated in Fig.~\ref{hvar}. Since the characteris features across all the phases are quite similar, here we consider only the case corresponding to the critical point, that is $\lvert t_1\rvert=t_2$. Three representative values of the chemical potentials are chosen, namely $\mu=-0.25,-0.5,$ and $-1$, denoted with their corresponding resutls with red, blue, and green colors, respectively. All the components $I_x$, $I_y$, and $I_z$ exhibits staircase-like pattern, just like in the $I_\alpha-\mu$ plots. The staircase structure arises because the spin persistent current is determined by the occupation of discrete eigenstates. As $h$ increases, the spin-dependent splitting modifies the energy spectrum. Each time an eigenvalue shifts across the chemical potential, the set of occupied states changes, leading to discontinuous jumps in the current.  The $x$-component exhibits the strongest response among the three as observed in Fig.~\ref{hvar}(a), reaching values up to $\sim 1.5\,$eV. The $y$-component (Fig.~\ref{hvar}(b)) is comparatively weaker, with magnitudes within $\pm 0.25\,$eV. The $z$-component lies in between the $x$ and $y$ responses in strength, with magnitudes up to $\sim 0.9\,$eV as shown in Fig.~\ref{hvar}(c). Two key observations emerge from the results: (i) the maximum magnitude of the spin persistent current occurs at lower $\mathpzc{h}$ values when the chemical potential is higher, and (ii) the current magnitude for all components gradually decreases as $\mathpzc{h}$ increases. These behaviors can be understood as follows. The chemical potential determines which energy levels are occupied, and hence which states contribute to the spin persistent current. When the chemical potential is closer to the band center (e.g., $\mu=-0.25$), many states near the Fermi level are active. This means even a small spin-dependent splitting $\mathpzc{h}$ is sufficient to generate strong contributions from these states, so the current reaches its maximum at lower $\mathpzc{h}$. As the chemical potential moves deeper into the band (e.g., $\mu=-0.5$ or $\mu=-1$), fewer states participate, requiring a larger $\mathpzc{h}$ to reach the peak current. In addition, increasing $\mathpzc{h}$ enhances the separation between up- and down-spin branches in the energy spectrum. This spin splitting reduces the overlap between spin channels and weakens the constructive interference that drives spin transport, causing the overall current magnitude to gradually decrease with larger $\mathpzc{h}$.

For completeness, we study the finite-size effect on the persistent spin currents by varying the number of unit cells from 10 to 60, while keeping the chemical potential fixed at $\mu = -0.5$. All other parameters are listed in the caption of Fig.~\ref{sysvar}, where we plot the three components of the persistent spin current as a function of the total number of sites ($2N$, $N$ being the number of unit cells). We consider three distinct regimes defined by the intradimer hopping amplitude $t_1$, with the $x$, $y$, and $z$ components of the persistent spin current shown in red, blue, and green, respectively. While the absolute magnitudes of the spin-current components depend on microscopic parameters such as the chemical potential, hopping strengths, SDS factor, and moment orientations, a systematic decrease of the spin current with increasing system size is observed in all regimes.

\begin{figure*}[ht] 
\includegraphics[width=0.3\textwidth]{fig15a.eps}\hfill
\includegraphics[width=0.3\textwidth]{fig15b.eps}\hfill
\includegraphics[width=0.3\textwidth]{fig15c.eps}
\caption{(Color online.) Im$\left[\text{BSBO}_z\right]$ as a function of the AAH modulation strength $W$. Panels correspond to (a) $\lvert t_1\rvert = 0.75$, (b) $\lvert t_1\rvert = 1$, and (c) $\lvert t_1\rvert = 1.25$. The system has $N = 20$ unit cells, with intercell hopping $t_2 = 1$ and magnetic flux $\Phi = 0.25$. The SDS parameter is $\mathpzc{h} = 0.5$, with $\theta = \pi/3$ and $\phi = \pi/9$. The chemical potential is set to $\mu = -0.5$.}
\label{bsbo}
\end{figure*}

Importantly, the rate of this decay exhibits a clear and robust phase dependence. In the topological phase (Fig.~\ref{sysvar}(a)), the decay is the fastest, and the spin current becomes negligibly small beyond $2N \approx 40$. At the critical point defined by $t_1=t_2$ (Fig.~\ref{sysvar}(b)), the spin current still decreases with system size. However, the decay is significantly slower, and all components remain finite even for the largest rings considered ($2N=120$). In the trivial phase (Fig.~\ref{sysvar}(c)), the decay is intermediate between these two limits. This finite-size trend is fully consistent with our previous work~\cite{ganguly2025}, where the persistent charge current exhibits a similar behavior with system dimension. 

Physically, the universal slowing of the decay at the critical point originates from the vanishing of dimerization at $t_1=t_2$, which allows spin-dependent hopping processes to propagate more uniformly around the ring. This enhances long-range biorthogonal spin-bond coherence and suppresses finite-size localization effects. Away from the critical point ($t_1\neq t_2$), dimerization selectively suppresses either intra- or intercell spin transport, leading to a more rapid decay of the spin current with increasing system size. Moreover, we find that the decay is steepest in the topological phase and moderate in the trivial phase. This can be understood from the fact that the effective hopping strength is larger in the trivial phase than in the topological phase, thereby supporting more extended spin transport and a slower decay compared to the topological regime.

Usually, disorder induces localization and consequently reduces the current response, a phenomenon popularly known as Anderson localization~\cite{anderson,gangof4}. However, the presence of disorder can have very interesting effects in non-Hermitian systems. For instance, in our earlier work~\cite{ganguly2025}, we observed disorder-induced amplification of the persistent charge current in an anti-Hermitian Hatano-Nelson ring. This is not an isolated example. For instance, in another recent study, directional amplification was found to persist even in the presence of disorder in a driven-dissipative cavity array with on-site disorder~\cite{cc-wan}. Similarly, other works have shown that disorder in a lattice can enhance transport~\cite{longi}. These findings make it highly worthwhile to investigate the effect of disorder on persistent spin currents.

To this end, we now introduce on-site energies $\epsilon_{n,\alpha}$, which were previously set to zero, via the well-known Aubry-Andr\'{e}-Harper (AAH) model~\cite{aah1,aah2,aah3}. For the $A$ sublattice, the on-site energy is given by $\epsilon_{n,A} = W \cos{[2\pi b (2n-1)]}$, and for the $B$ sublattice, $\epsilon_{n,B} = W \cos{[2\pi b (2n)]}$. Here, $W$ denotes the disorder strength, and $b$ is an irrational number taken as $b = (1+\sqrt{5})/2$. The results are shown in Fig.~\ref{disvar}, where we plot the three components of the persistent spin current as a function of the AAH modulation strength $W$ across the three previously considered regimes. The chemical potential is fixed at $\mu = -0.5$, and all other parameters are listed in the caption of Fig.~\ref{disvar}. Interestingly, amplification occurs in all cases. In the topological phase (Fig.~\ref{disvar}(a)), the currents are vanishingly small for small $W$, as shown in the inset for $W$ between 0 and 0.5. Beyond $W = 0.5$, amplification sets in, reaching a maximum about $0.6\,$eV for $I_x$ around $W \approx 0.8$. For $W \approx 1$, localization dominates, and the currents again become negligible. At the critical point (Fig.~\ref{disvar}(b)), the currents initially decrease with increasing $W$, with a maximum magnitude of about $1.2\,$eV for $I_x$, then begin to increase near $W = 1$, reaching a peak of about $0.6\,$eV, indicating amplification. All currents vanish again beyond $W \approx 1.25$, following the same trend with differing magnitudes. In the trivial phase (Fig.~\ref{disvar}(c)), the currents increase monotonically with $W$, reaching a peak around $W \approx 0.8$, where $I_x$ attains the highest magnitude among all cases, about $6\,$eV, before decreasing toward zero. A small secondary peak is observed near $W \approx 1.25$, but for larger $W$, localization dominates and all currents vanish.

To elucidate the microscopic origin of the disorder dependence of the spin current in the non-Hermitian ferromagnetic HN ring, we analyze a bond-resolved biorthogonal quantity that captures spin-dependent coherence between neighboring lattice sites. In non-Hermitian systems, expectation values of physical observables must be evaluated using biorthogonal left and right eigenstates, and the resulting correlators are, in general, complex quantities~\cite{biortho,kawabata}. For that, we define the biorthogonal spin-bond overlap (BSBO) on a nearest-neighbor bond $i\rightarrow i+1$ as
\begin{equation}
\text{BSBO}^{\alpha,(n)}_{i,i+1} = \langle\psi_n^L\vert\bm{c}_i^\dagger\bm{\sigma}_\alpha \bm{c}_{i+1}\vert\psi_n^R\rangle,\quad\quad \alpha=x,y,z,
\end{equation}
where $\vert\psi_n^L\rangle$ and $\langle\psi_n^R\vert$ are the left and right eigenstates of the non-Hermitian Hamiltonian, $\bm{c}_i^\dagger$ and $\bm{c}_j$ are the creation and annihilation operators, respectively as defined in Eq.~\ref{ham}, and $\bm{\sigma}_\alpha$ are the Pauli matrices acting in spin space.

For each occupied eigenstate $n$, summing over all bonds yields the spin-bond overlap per eigenstate
\begin{equation}
\text{BSBO}^{\alpha,(n)} = \sum_{\langle i,i+1\rangle} \text{BSBO}^{\alpha,(n)}_{i,i+1}.
\end{equation}

Finally, we compute the average BSBO by averaging the state-resolved BSBO over all occupied states with real part of the eigenenergy below the chemical potential $\mu$, which reads as
\begin{equation}
\text{BSBO}_\alpha = \dfrac{1}{N_{\text{occ}}}\sum_{\text{Re}\left(E_n\right)\leq \mu} \text{BSBO}^{\alpha,(n)},
\end{equation}
where $N_{\text{occ}}$ denotes the number of occupied states.

The BSBO is generally complex. However, only its imaginary part contributes to the spin current. This follows from the anti-Hermitian structure of the lattice current operator, which is obtained either from the continuity equation or equivalently as the derivative of the Hamiltonian with respect to the Peierls phase. As a result, the current selectively probes the phase-dependent component of the biorthogonal bond correlator, while the real part reflects amplitude-like bond coherence that does not contribute to net transport~\cite{kawabata}.

Since the disorder dependence of the three spin-current components is qualitatively similar apart from their magnitudes, we focus on the imaginary part of the $z$-component of the biorthogonal spin-bond overlap and plot Im$\left[\text{BSBO}_z\right]$ as a function of disorder strength in Fig.~\ref{bsbo}. We find that Im$\left[\text{BSBO}_z\right]$ closely follows the disorder dependence of the spin current in all three regimes (topological, critical, and trivial), indicating that it provides a microscopic measure of the spin-dependent phase coherence governing transport in the non-Hermitian ferromagnetic HN ring.

The nonmonotonic disorder dependence of the spin current observed across the three regimes is therefore closely connected to the behavior of Im$\left[\text{BSBO}_z\right]$. It is important to note that for the AAH model with isotropic nearest-neighbor hopping in one dimension, a sharp delocalization-to-localization transition occurs at $W = 2t$. Consequently, within the disorder window considered in Figs.~\ref{disvar} and \ref{bsbo}, the system predominantly resides in the extended regime. Nevertheless, the BSBO profiles indicate that disorder plays a dual role: at moderate strength, it enhances the biorthogonal spin-bond overlap and thereby promotes spin transport, whereas at stronger disorder it suppresses Im$\left[\text{BSBO}_z\right]$ due to the gradual localization of eigenstates. The distinct disorder responses observed in the three regimes thus reflect differences in how biorthogonal phase coherence is reorganized by disorder, rather than signaling sharp transitions between extended and localized states.

\section{\label{conclusion}Summary}
To summarize, we have investigated the behavior of persistent charge and spin currents in a ferromagnetic Hatano-Nelson ring. The non-reciprocal intradimer hopping acts as a synthetic magnetic flux, giving rise to a non-Hermitian Aharonov-Bohm effect, while ferromagnetic ordering generates spin splitting. The system is modeled using a nearest-neighbor tight-binding framework. We derived the analytical form of the dispersion relation and analyzed the corresponding complex band structure. Persistent charge and all three components of the spin current were computed using the current operator in the biorthogonal basis, with a detailed mathematical derivation provided. We examined how various factors, including the orientation of magnetic moments, chemical potential, spin-dependent scattering parameter, finite-size effects, and disorder, affect the currents. The key findings of our work are mentioned point by point below.

$\bullet$ The ferromagnetic exchange field splits up- and down-spin bands in the real spectrum, while the imaginary spectrum remains spin-degenerate, simply shifting the energy window without altering spectral features.

$\bullet$ The intra- to interdimer hopping ratio controls the bonding-antibonding separation. The observation for $h=0$ with a gapped real spectrum in the topological phase, gapless at the critical point, and a gapped imaginary spectrum in the trivial phase remains valid for each spin separately when $\mathpzc{h}\neq 0$, with $\mathpzc{h}$ simply shifting the up- and down-spin bands without altering the bonding-antibonding structure.

$\bullet$ The real and imaginary persistent charge currents are no longer identical in the presence of the ferromagnetic exchange field at the critical point.

$\bullet$ Real persistent spin currents do not follow the same phase-dependent trends as charge currents, while the imaginary spin currents always vanish since there is no spin splitting in the imaginary spectrum.

$\bullet$ The components respond differently to the orientation of magnetic moments. $I_z$ is antisymmetric in the polar angle ($\theta$) and independent of the azimuthal angle ($\varphi$), while $I_x$ and $I_y$ are symmetric in $\theta$ and vary as $\cos{\varphi}$ and $\sin{\varphi}$, respectively.

$\bullet$ Persistent spin currents exhibit a staircase-like pattern due to the discrete energy spectrum and are antisymmetric about $\mu=0$ because of particle-hole symmetry.

$\bullet$ As $\mathpzc{h}$ increases, spin splitting reduces overlap between spin channels, weakening constructive interference and gradually reducing current magnitudes.

$\bullet$ Lower chemical potentials require smaller $\mathpzc{h}$ to achieve maximum spin current.

$\bullet$ The magnitudes of all spin-current components decrease with increasing system size, decaying fastest in the topological phase, slowest at the critical point, and moderately in the trivial phase, with trends mirroring those of persistent charge currents and indicating similar dimensional scaling across spin and charge channels.

$\bullet$ Moderate AAH-type disorder can strongly enhance persistent spin currents, with peak magnitudes far exceeding the clean system, before localization eventually suppresses the currents at higher disorder strengths.

These results not only reveal the rich interplay of non-Hermiticity, ferromagnetic ordering, and disorder in controlling charge and spin currents, but also demonstrate a new framework for analyzing spin-dependent persistent currents in non-Hermitian systems. To our knowledge, the spin-current formulation presented here is entirely new, providing a foundation for future theoretical and experimental studies of spin transport in non-Hermitian topological systems.

Finally, we note that in the present model the persistent spin current is purely real, reflecting the absence of spin splitting in the imaginary spectrum. It would be interesting to explore more general non-Hermitian settings, such as spin-selective gain and loss or non-reciprocal spin-dependent processes, in which spin-dependent non-Hermiticity may generate an imaginary spin current, an issue left for future work.

\appendix

\section{Biorthogonalization of Eigenstates}
\label{app:biorthogal}
In non-Hermitian systems, the eigenvalue problem is defined through the right and left eigenstates,
\begin{equation}
    H \lvert\psi^R_n\rangle = E_n \lvert\psi^R_n\rangle, 
    \qquad 
    H^\dagger \langle\psi^L_n\rvert = E_n^* \langle\psi^L_n\rvert,
\end{equation}
where $\lvert\psi^R_n\rangle$ and $\langle\psi^L_n\rvert$ are the right and left eigenvectors associated with the eigenvalue $E_n$, respectively. Unlike the Hermitian case, right eigenvectors alone do not form an orthogonal basis. Instead, a consistent framework requires the introduction of a {\it biorthogonal basis}~\cite{biortho}, where the left and right eigenstates satisfy the condition
\begin{equation}
    \langle\psi^L_m\rvert \psi^R_n\rangle = \delta_{mn}.
\end{equation}

Below we provide the procedures to get the biorthogonal basis efficiently. 
\subsection{Non-degenerate case} 
When the eigenvalues are non-degenerate, the procedure reduces to a straightforward normalization of each left-right eigenpair. For a given eigenstate, the overlap is
\begin{equation}
N_n = \langle\psi^L_n\rvert \psi^R_n\rangle.
\end{equation}
The biorthonormalized eigenvectors are defined as
\begin{equation}
\rvert \psi^R_n\rangle \rightarrow \frac{\psi^R_n\rangle}{\sqrt{N_n}},\quad \langle\psi^L_n\rvert\rightarrow \frac{\langle\psi^L_n\rvert}{\sqrt{N_n^*}},
\end{equation}
so that one gets 
\begin{equation}
\langle\psi^L_n\rvert \psi^R_n\rangle = 1.
\label{normal}
\end{equation}
It is to be be noted that since the eigenvalues are non-degenerate, eigenvectors associated with different eigenvalues are automatically biorthogonal, that is
\begin{equation}
\langle\psi^L_m\rvert \psi^R_n\rangle = 0, \quad m\neq n.
\label{orthogonal}
\end{equation} 
Therefore, with Eqs.~\ref{normal} and \ref{orthogonal}, one can construct the biorthogonal basis for non-degenerate case.

\subsection{Degenerate case} 
If the Hamiltonian hosts degenerate eigenvalues, the associated left and right eigenvectors span degenerate subspaces. In such cases, a more careful procedure is required to establish a biorthogonal basis. A common approach is to employ a \emph{Gram-Schmidt-type biorthogonalization}~\cite{biortho1,biortho2}. Suppose $R_{\rm block}$ and $L_{\rm block}$ denote the matrices whose columns are the right and left eigenvectors corresponding to a degenerate eigenvalue. The overlap matrix between them is given by
\begin{equation}
    B = L_{\rm block}^\dagger R_{\rm block}.
\end{equation}
To enforce biorthogonality, one defines
\begin{equation}
    R_{\rm ortho} = R_{\rm block}\, B^{-1/2}, 
    \quad 
    L_{\rm ortho} = L_{\rm block}\, \big(B^{-1/2}\big)^\dagger,
\end{equation}
where $B^{-1/2}$ denotes the inverse square root of $B$. This transformation guarantees
\begin{equation}
    L_{\rm ortho}^\dagger R_{\rm ortho} = \mathbb{I},
\end{equation}
so that the biorthogonality condition is satisfied within the degenerate subspace. This construction is analogous to block-orthogonalization procedures in numerical linear algebra, but adapted for left-right eigenpairs in non-Hermitian systems.

Such biorthogonalization is essential in non-Hermitian quantum mechanics and condensed matter physics, since physical observables and correlation functions are naturally expressed in terms of both left and right eigenstates. In particular, quantities like Berry curvature, response coefficients, and persistent currents are well-defined only once a properly normalized biorthogonal basis is established.

\section{Derivation of the charge current}
\label{app:current}
Starting from the definition
\begin{equation}
\hat{J} = \frac{e\dot{X}}{2Na}, \qquad 
\dot{X} = \frac{2\pi}{ih}[X,H],
\end{equation}
with the position operator
\begin{equation}
X = \sum_{n=1}^N n a \, \left(\bm{c}_{n,A}^\dagger \bm{c}_{n,A} + \bm{c}_{n,B}^\dagger \bm{c}_{n,B}\right),
\end{equation}
we evaluate the commutator $[X,H]$ for the present Hamiltonian.  
This yields the current operator

\begin{eqnarray}
\hat{J} &=& -\frac{2\pi e |t_1|}{i 2N h} 
\sum_{n=1}^N \left( e^{i\phi} c_{n,A\uparrow}^\dagger c_{n,B\uparrow} 
+ e^{-i\phi} c_{n,B\uparrow}^\dagger c_{n,A\uparrow} \right) \nonumber\\
&-& \frac{2\pi e |t_1|}{i 2N h} 
\sum_{n=1}^N \left( e^{i\phi} c_{n,A\downarrow}^\dagger c_{n,B\downarrow} 
+ e^{-i\phi} c_{n,B\downarrow}^\dagger c_{n,A\downarrow} \right) \nonumber\\
&-& \frac{2\pi e t_2}{i 2N h} 
\sum_{n=1}^{N-1} \left( c_{n,B\uparrow}^\dagger c_{n+1,A\uparrow} 
- c_{n+1,A\uparrow}^\dagger c_{n,B\uparrow} \right) \nonumber\\
&-& \frac{2\pi e t_2}{i 2N h} 
\sum_{n=1}^{N-1} \left( c_{n,B\downarrow}^\dagger c_{n+1,A\downarrow} 
- c_{n+1,A\downarrow}^\dagger c_{n,B\downarrow} \right) \nonumber\\
&+& \frac{2\pi e t_2}{i 2N h} 
\left( c_{N,B\uparrow}^\dagger c_{1,A\uparrow} - c_{1,A\uparrow}^\dagger c_{N,B\uparrow} \right) \nonumber\\
&+& \frac{2\pi e t_2}{i 2N h} 
\left( c_{N,B\downarrow}^\dagger c_{1,A\downarrow} - c_{1,A\downarrow}^\dagger c_{N,B\downarrow} \right).
\end{eqnarray}

The persistent charge current for the $n$th eigenstate can be determined as follows
\begin{equation}
I^n = \left\langle \psi_n^{L} \middle| \hat{J}\middle| \psi_n^{R} \right\rangle,
\end{equation}
where $\left| \psi_n^{L} \right\rangle$ and $\left| \psi_n^{R} \right\rangle$ are the left and right eigenvectors, respectively.

The right eigenstate $\left|\psi_n^{R}\right\rangle$ can be written as
\begin{equation}
\left|\psi_n^{R}\right\rangle = \sum_{p}\sum_{\alpha=A,B} \left( a_{p, \alpha\uparrow}^{n,R} \, |p , \alpha\uparrow\rangle + a_{p , \alpha\downarrow}^{n,R} \, |p , \alpha\downarrow\rangle \right).
\end{equation}
Here, $\lvert p, \alpha \uparrow \rangle$ and $\lvert p, \alpha \downarrow \rangle$ denote the Wannier states at sublattice $\alpha$ with spin up and spin down, respectively. The quantities $a_{p, \alpha \uparrow}^{n,R}$ and $a_{p, \alpha \downarrow}^{n,R}$ are the corresponding expansion coefficients of the right eigenstate $\lvert \psi_n^{R} \rangle$.

Similarly the left eigenstate $\left|\psi_n^{L}\right\rangle$ can be written as
\begin{equation}
\left|\psi_n^{L}\right\rangle = \sum_{p}\sum_{\alpha=A,B} \left( a_{p, \alpha\uparrow}^{n,L} \, |p , \alpha\uparrow\rangle + a_{p , \alpha\downarrow}^{n,L} \, |p , \alpha\downarrow\rangle \right).
\end{equation} 
$a_{p \uparrow}^{n,L}$ and $a_{p \downarrow}^{n,L}$ are the corresponding coefficients for the left eigenstate $\left| \psi_n^{L} \right\rangle$.
The current for the $n$th eigenstate takes the form
\begin{align}
I^n &= -\frac{2\pi e |t_1|}{i 2N h}
\sum_{n=1}^N \left( e^{i\phi}\left(a_{n,A\uparrow}^{L}\right)^{*} a_{n,B\uparrow}^{R}
+ e^{-i\phi}\left(a_{n,B\uparrow}^{L}\right)^{*} a_{n,A\uparrow}^{R} \right) \notag\\
&\quad -\frac{2\pi e |t_1|}{i 2N h}
\sum_{n=1}^N \left( e^{i\phi}\left(a_{n,A\downarrow}^{L}\right)^{*} a_{n,B\downarrow}^{R}
+ e^{-i\phi}\left(a_{n,B\downarrow}^{L}\right)^{*} a_{n,A\downarrow}^{R} \right) \notag\\
&\quad -\frac{2\pi e t_2}{i 2N h}
\sum_{n=1}^{N-1} \left( \left(a_{n,B\uparrow}^{L}\right)^{*} a_{n+1,A\uparrow}^{R}
- \left(a_{n+1,A\uparrow}^{L}\right)^{*} a_{n,B\uparrow}^{R} \right) \notag\\
&\quad -\frac{2\pi e t_2}{i 2N h}
\sum_{n=1}^{N-1} \left( \left(a_{n,B\downarrow}^{L}\right)^{*} a_{n+1,A\downarrow}^{R}
- \left(a_{n+1,A\downarrow}^{L}\right)^{*} a_{n,B\downarrow}^{R} \right) \notag\\
&\quad +\frac{2\pi e t_2}{i 2N h}
\left( \left(a_{N,B\uparrow}^{L}\right)^{*} a_{1,A\uparrow}^{R} - \left(a_{1,A\uparrow}^{L}\right)^{*} a_{N,B\uparrow}^{R} \right) \notag\\
&\quad +\frac{2\pi e t_2}{i 2N h}
\left( \left(a_{N,B\downarrow}^{L}\right)^{*} a_{1,A\downarrow}^{R} - \left(a_{1,A\downarrow}^{L}\right)^{*} a_{N,B\downarrow}^{R} \right).
\label{charge-curr}
\end{align}

\section{Derivation of spin current components}
\label{app:spin_current}
The spin current operator is defined as
\begin{equation}
\hat{I}_\alpha = \frac{\hbar}{4aN} \left( \bm{\sigma}_\alpha \dot{X} + \dot{X}\,\bm{\sigma}_\alpha \right),
\quad \alpha \in \{x,y,z\},
\end{equation}
where $\sigma_\alpha$ are the Pauli matrices. 

\subsection{$x$-component}
The expression for the $x$-component of the spin current operator is given by
\begin{equation}
\hat{I}_x=\frac{\hbar\left(\sigma_x \dot{X}+\dot{X} \sigma_x\right)}{4 a N}.
\end{equation}

By substituting $\sigma_x$ and $\dot{X}$ into the equation and doing some algebra similar to charge current, we derive a closed form for the spin current operator. 

The operator form of the $x$-component of the spin current is given by
\begin{align}
\hat{I}_x = &- \frac{2\pi|t_1|}{2Ni}\sum_{n=1}^N \left[ 
 e^{i \phi} \left( c_{n,A\uparrow}^\dagger c_{n,B\downarrow} + c_{n,A\downarrow}^\dagger c_{n,B\uparrow} \right) \right.\notag \\
&\left. +  e^{-i \phi} \left( c_{n,B\uparrow}^\dagger c_{n,A\downarrow} + c_{n,B\downarrow}^\dagger c_{n,A\uparrow} \right) \right] \notag \\
 &+ \frac{2\pi t_2}{2Ni}\sum_{n = 1}^{N-1} \left[ \left( c_{n+1,B\uparrow}^\dagger c_{n,A\downarrow} + c_{n+1,B\downarrow}^\dagger c_{n,A\uparrow} \right) \right. \notag \\
&\left. -\left( c_{n,A\uparrow}^\dagger c_{n+1,B\downarrow} + c_{n,A\downarrow}^\dagger c_{n+1,B\uparrow} \right)\right] \notag \\
 &+  \frac{2\pi t_2}{2Ni} \left[\left( c_{N,B\uparrow}^\dagger c_{1,A\downarrow} + c_{N,B\downarrow}^\dagger c_{1,A\uparrow} \right)\right.\notag\\
 &\left.- \left( c_{1,A\uparrow}^\dagger c_{N,B\downarrow} + c_{1,A\downarrow}^\dagger c_{N,B\uparrow} \right)\right]
\end{align}

Using the operation $\left\langle \psi_n^{L} \middle| \hat{I}_{x} \middle| \psi_n^{R} \right\rangle$ the $x$-component of the current for the $n$th eigenstate takes the form
\begin{align}
I_x^n = &- \frac{2\pi|t_1|}{2Ni} \sum_{n=1}^N \left[
 e^{i\phi} \left\{ \left(a_{n,A\uparrow}^{L}\right)^{*} a_{n,B\downarrow}^{R}
+ \left(a_{n,A\downarrow}^{L}\right)^{*} a_{n,B\uparrow}^{R} \right\}\right. \notag\\
&\left.+ e^{-i\phi} \left\{\left(a_{n,B\uparrow}^{L}\right)^{*} a_{n,A\downarrow}^{R}
+ \left(a_{n,B\downarrow}^{L}\right)^{*} a_{n,A\uparrow}^{R} \right\}
\right] \notag\\
& + \frac{2\pi t_2}{2Ni}\sum_{n=1}^{N-1} \left[
\left\{ \left(a_{n+1,B\uparrow}^{L}\right)^{*} a_{n,A\downarrow}^{R}
+ \left(a_{n+1,B\downarrow}^{L}\right)^{*} a_{n,A\uparrow}^{R} \right\}\right.\notag\\
&\left.-\left\{ \left(a_{n,A\uparrow}^{L}\right)^{*} a_{n+1,B\downarrow}^{R}
+ \left(a_{n,A\downarrow}^{L}\right)^{*} a_{n+1,B\uparrow}^{R} \right\}
\right] \notag\\
& + \frac{2\pi t_2}{2Ni}\left[ \left\{\left(a_{N,B\uparrow}^{L}\right)^{*} a_{1,A\downarrow}^{R}
+ \left(a_{N,B\downarrow}^{L}\right)^{*} a_{1,A\uparrow}^{R} \right\} \right.\notag\\
&\left. - \left\{ \left(a_{1,A\uparrow}^{L}\right)^{*} a_{N,B\downarrow}^{R}
+ \left(a_{1,A\downarrow}^{L}\right)^{*} a_{N,B\uparrow}^{R} \right\}\right].
\end{align}

\subsection{$y$-component}
The spin current operator corresponding to the $y$-component is defined as
\begin{equation}
\hat{I}_y=\frac{\hbar\left(\sigma_y \dot{X}+\dot{X} \sigma_y\right)}{4 a N}.
\end{equation}

By substituting $\sigma_y$ and $\dot{X}$ into the equation and doing similar algebra like the $x$-comopnent, we get the operator form for the $y$-component of the spin current operator. 

The $y$-component of the spin current operator can be expressed as
\begin{align}
\hat{I}_y = &- \frac{2\pi|t_1|}{2N} \sum_{n=1}^N \left[
 e^{i\phi}\left( -c_{n,A\uparrow}^\dagger c_{n,B\downarrow} + c_{n,A\downarrow}^\dagger c_{n,B\uparrow} \right)\right. \notag\\
&\left.+ e^{-i\phi}\left( - c_{n,B\uparrow}^\dagger c_{n,A\downarrow} + c_{n,B\downarrow}^\dagger c_{n,A\uparrow} \right)
\right] \notag\\
& + \frac{2\pi t_2}{2N}\sum_{n = 1}^{N-1} \left[
\left( -c_{n+1,B\uparrow}^\dagger c_{n,A\downarrow} + c_{n+1,B\downarrow}^\dagger c_{n,A\uparrow} \right) \right. \notag\\
&
\left. -\left( - c_{n,A\uparrow}^\dagger c_{n+1,B\downarrow} + c_{n,A\downarrow}^\dagger c_{n+1,B\uparrow} \right)\right] \notag\\
& +\frac{2\pi t_2}{2N}\left[ \left(- c_{N,B\uparrow}^\dagger c_{1,A\downarrow} + c_{N,B\downarrow}^\dagger c_{1,A\uparrow} \right)\right. \notag\\
&\left.-\left( - c_{1,A\uparrow}^\dagger c_{N,B\downarrow} + c_{1,A\downarrow}^\dagger c_{N,B\uparrow} \right)\right].
\end{align}

The $y$-component of the spin current for the $n$th eigenstate is obtained by evaluating the matrix element $\left\langle \psi_n^{L} \middle| \hat{I}_y \middle| \psi_n^{R} \right\rangle$
\begin{align}
I_y^n = &- \frac{2\pi|t_1|}{2N}\sum_{n=1}^N \left[
 e^{i\phi}\,\left\{
- \left(a_{n,A\uparrow}^{L}\right)^{*} a_{n,B\downarrow}^{R} + \left(a_{n,A\downarrow}^{L}\right)^{*} a_{n,B\uparrow}^{R} \right\} \right.\notag\\
&\left.
+ e^{-i\phi} \left\{ 
- \left(a_{n,B\uparrow}^{L}\right)^{*} a_{n,A\downarrow}^{R} + \left(a_{n,B\downarrow}^{L}\right)^{*} a_{n,A\uparrow}^{R} \right\}
\right] \notag\\
&+ \frac{2\pi t_2}{2N}\sum_{n=1}^{N-1} \left[
\left\{ 
- \left(a_{n+1,B\uparrow}^{L}\right)^{*} a_{n,A\downarrow}^{R} + \left(a_{n+1,B\downarrow}^{L}\right)^{*} a_{n,A\uparrow}^{R} \right\}\right. \notag\\
&\left.
-\left\{ 
- \left(a_{n,A\uparrow}^{L}\right)^{*} a_{n+1,B\downarrow}^{R} + \left(a_{n,A\downarrow}^{L}\right)^{*} a_{n+1,B\uparrow}^{R} \right\}
\right] \notag\\
& + \frac{2\pi t_2}{2N}\left[ \left\{
- \left(a_{N,B\uparrow}^{L}\right)^{*} a_{1,A\downarrow}^{R} + \left(a_{N,B\downarrow}^{L}\right)^{*} a_{1,A\uparrow}^{R} \right\}\right. \notag\\
&\left. - \left\{ 
- \left(a_{1,A\uparrow}^{L}\right)^{*} a_{N,B\downarrow}^{R} + \left(a_{1,A\downarrow}^{L}\right)^{*} a_{N,B\uparrow}^{R} \right\}\right].
\end{align}

\subsection{$z$-component}
The spin current operator associated with the $z$-component is defined as
\begin{equation}
\hat{I}_z=\frac{\hbar\left(\sigma_z \dot{X}+\dot{X} \sigma_z\right)}{4 a N}.
\end{equation}

By plugging $\sigma_z$ and $\dot{X}$ into the equation and carrying out algebraic steps similar to those for the $x$-component, we arrive at the operator form of the $z$-component of the spin current. The operator corresponding to the $z$-component of the spin current is expressed as
\begin{align}
\hat{I}_z = &- \frac{2\pi|t_1|}{2Ni}\sum_{n=1}^N \left[ 
e^{i \phi} \left( c_{n,A\uparrow}^\dagger c_{n,B\uparrow} - c_{n,A\downarrow}^\dagger c_{n,B\downarrow} \right) \right.\notag \\
& \left.+ e^{-i \phi} \left( c_{n,B\uparrow}^\dagger c_{n,A\uparrow} - c_{n,B\downarrow}^\dagger c_{n,A\downarrow} \right) \right] \notag \\
& + \frac{2\pi t_2}{2Ni}\sum_{n = 1}^{N-1} \left[ 
\left( c_{n+1,B\uparrow}^\dagger c_{n,A\uparrow} - c_{n+1,B\downarrow}^\dagger c_{n,A\downarrow} \right) \right.\notag \\
&\left. -  \left( c_{n,A\uparrow}^\dagger c_{n+1,B\uparrow} - c_{n,A\downarrow}^\dagger c_{n+1,B\downarrow} \right) \right] \notag \\
& + \frac{2\pi t_2}{2Ni} \left[\left( c_{N,B\uparrow}^\dagger c_{1,A\uparrow} - c_{N,B\downarrow}^\dagger c_{1,A\downarrow} \right)\right. \notag \\
&\left.-  \left( c_{1,A\uparrow}^\dagger c_{N,B\uparrow} - c_{1,A\downarrow}^\dagger c_{N,B\downarrow} \right)\right].
\end{align}

Applying the operation $\left\langle \psi_n^{L} \middle| \hat{I}_z \middle| \psi_n^{R} \right\rangle$, the $z$-component of the spin current for the $n$th eigenstate is expressed as follows
\begin{align}
I_z^n
= &- \frac{2\pi|t_1|}{2Ni}\sum_{n=1}^N \left[ 
 e^{i \phi} \left\{ \left(a_{n,A\uparrow}^{L}\right)^{*} a_{n,B\uparrow}^{R} - \left(a_{n,A\downarrow}^{L}\right)^{*} a_{n,B\downarrow}^{R} \right\}\right. \notag \\
& \left. + e^{-i \phi} \left\{ \left(a_{n,B\uparrow}^{L}\right)^{*} a_{n,A\uparrow}^{R} - \left(a_{n,B\downarrow}^{L}\right)^{*} a_{n,A\downarrow}^{R} \right\} \right] \notag \\
& + \frac{2\pi t_2}{2Ni}\sum_{n = 1}^{N-1} \left[ 
 \left\{ \left(a_{n+1,B\uparrow}^{L}\right)^{*} a_{n,A\uparrow}^{R} - \left(a_{n+1,B\downarrow}^{L}\right)^{*} a_{n,A\downarrow}^{R} \right\}\right. \notag \\
& \left. -  \left\{ \left(a_{n,A\uparrow}^{L}\right)^{*} a_{n+1,B\uparrow}^{R} - \left(a_{n,A\downarrow}^{L}\right)^{*} a_{n+1,B\downarrow}^{R} \right\} \right] \notag \\
& + \frac{2\pi t_2}{2Ni} \left[ \left\{\left(a_{N,B\uparrow}^{L}\right)^{*} a_{1,A\uparrow}^{R} - \left(a_{N,B\downarrow}^{L}\right)^{*} a_{1,A\downarrow}^{R} \right\}\right. \notag \\
&\left. -  \left\{ \left(a_{1,A\uparrow}^{L}\right)^{*} a_{N,B\uparrow}^{R} - \left(a_{1,A\downarrow}^{L}\right)^{*} a_{N,B\downarrow}^{R} \right\}\right].
\end{align}

\section{Derivation of Bloch Hamiltonian and dispersion relations}
\label{app:kspace}
Through the Fourier transformation, the creation and annihilation operators can be written as
\begin{eqnarray}
c_{n,\alpha\sigma} &=& \frac{1}{\sqrt{2N}} \sum_k e^{ikna} c_{k,\alpha\sigma},\\
c_{n,\alpha\sigma}^\dagger &=& \frac{1}{\sqrt{2N}} \sum_k e^{-ikna} c_{k,\alpha\sigma}^\dagger,
\end{eqnarray}
where $k=\frac{2\pi m}{2N}$ with $m=0,1,\ldots,2N-1$ for periodic boundary conditions. Here, $n$ denotes the unit-cell index, $\alpha\in{A,B}$ labels the sublattice, and $\sigma\in{\uparrow,\downarrow}$ the spin.

Substitute the Fourier transforms into the real-space Hamiltonian as given in Eq.~\ref{ham} and following the standard procedure, the Hamiltonian in $k$-space in the basis $[A\uparrow, A\downarrow, B\uparrow, B\downarrow]$ becomes
\begin{widetext}
$H(k) =
\begin{pmatrix}
\epsilon - \mathpzc{h}\cos{\theta} & -\mathpzc{h}\sin{\theta}e^{-i\varphi} & \lvert t_1 \rvert e^{i\phi}+t_2 e^{-ik} & 0 \\
-\mathpzc{h}\sin{\theta} e^{i\varphi} & \epsilon + \mathpzc{h}\cos{\theta} & 0 & \lvert t_1 \rvert e^{i\phi}+t_2 e^{-ik} \\
-\lvert t_1 \rvert e^{-i\phi}+t_2 e^{ik} & 0 & \epsilon - \mathpzc{h}\cos{\theta}  & -\mathpzc{h}\sin{\theta}e^{-i\varphi} \\
0 & -\lvert t_1 \rvert e^{-i\phi}+t_2 e^{ik} & -\mathpzc{h}\sin{\theta}e^{-i\varphi} & \epsilon + \mathpzc{h}\cos{\theta}
\end{pmatrix}.$
\end{widetext}
The $k$-space Hamiltonian is obtained under the assumptions $\epsilon_{n,\alpha}=\epsilon$ and $\mathpzc{h}_n=\mathpzc{h}$ for all $n$, $\theta_n=\theta$, and $\varphi_n=\varphi$.
The eigenvalues of $H(k)$ can be obtained as follows.

We write the Bloch Hamiltonian in block form as
\begin{align}
H(k) =
\begin{pmatrix}
A & B\mathbb{I}_2 \\
B^{\prime} \mathbb{I}_2 & A
\end{pmatrix},
\end{align}
where $A$ operates in spin space, and $B,B^\prime$ represent scalar hopping amplitudes that couple the sublattice degrees of freedom through the spin identity ${\mathbb I}_2$, and
\begin{align}
B &= |t_1| e^{i\phi} + t_2 e^{-ik}, \\
B^{\prime} &= -|t_1| e^{-i\phi} + t_2 e^{ik}, \\
A &=
\begin{pmatrix}
\epsilon - \mathpzc{h}\cos\theta & -\mathpzc{h}\sin\theta, e^{-i\varphi} \\
\mathpzc{h}\sin\theta, e^{i\varphi} & \epsilon + \mathpzc{h}\cos\theta
\end{pmatrix}.
\end{align}

For convenience, we define the matrix acting in sublattice space as
\begin{align}
D =
\begin{pmatrix}
0 & B \\
B^{\prime} & 0
\end{pmatrix}.
\end{align}
Since $A$ and $D$ operate in spin and sublattice spaces, respectively, they commute.
The Hamiltonian can therefore be written in a compact tensor-product form as
\begin{align}
H(k) = \mathbb{I}_2 \otimes A + D \otimes \mathbb{I}_2 ,
\end{align}
which facilitates diagonalization and clarifies the separable structure between spin and sublattice degrees of freedom.

The eigenvalues of $A$ are
\begin{align}
\lambda_A = \epsilon \pm \mathpzc{h},
\end{align}
and the eigenvalues of $D$ are
\begin{align}
\lambda_D = \pm \sqrt{B B^\prime} ,
\end{align}
with
\begin{align}
B B^\prime &= (\lvert t_1 \rvert e^{i\phi} + t_2 e^{-i k})(-\lvert t_1 \rvert e^{-i\phi} + t_2 e^{i k}) \\
&= t_2^2 - \lvert t_1 \rvert^2 + 2 i \lvert t_1 \rvert t_2 \sin(\phi + k).
\end{align}

Since $A$ and $D$ commute, the eigenvalues of $H(k)$ are simply additive, giving four bands:
\begin{align}
E &= (\epsilon \pm \mathpzc{h}) \pm \sqrt{t_2^2 - \lvert t_1 \rvert^2 + 2 i \lvert t_1 \rvert t_2 \sin(\phi + k)} .
\end{align}

Explicitly,
\begin{align}
E_1 &= \epsilon + \mathpzc{h} + \sqrt{t_2^2 - \lvert t_1 \rvert^2 + 2 i \lvert t_1 \rvert t_2 \sin(\phi + k)}, \\
E_2 &= \epsilon + \mathpzc{h} - \sqrt{t_2^2 - \lvert t_1 \rvert^2 + 2 i \lvert t_1 \rvert t_2 \sin(\phi + k)}, \\
E_3 &= \epsilon - \mathpzc{h} + \sqrt{t_2^2 - \lvert t_1 \rvert^2 + 2 i \lvert t_1 \rvert t_2 \sin(\phi + k)}, \\
E_4 &= \epsilon - \mathpzc{h} - \sqrt{t_2^2 - \lvert t_1 \rvert^2 + 2 i \lvert t_1 \rvert t_2 \sin(\phi + k)}.
\end{align}

Compactly, the eigenvalues can be written as
\begin{equation}
E_{s,\alpha}(k+\phi) = \epsilon + s\,\mathpzc{h} + \alpha \sqrt{t_2^2 - \lvert t_1 \rvert^2 + 2 i \lvert t_1 \rvert t_2 \sin(k+\phi)},
\label{dispers1}
\end{equation}
where $\sigma = +1$ ($\uparrow$ spin) and $\sigma = -1$ ($\downarrow$ spin). 
The upper sign $(+)$ corresponds to the antibonding band, while the lower sign $(-)$ corresponds to the bonding band.

The nomenclature follows directly from the eigenvalue expression. The square-root term in Eq.~\ref{dispers1} originates from the sublattice off-diagonal coupling and hence reflects the bonding-antibonding splitting, with the $+$ sign denoting the antibonding branch and the $-$ sign the bonding branch. On the other hand, the $\pm \mathpzc{h}$ shift arises from the spin-dependent exchange field in the term $A$, which rigidly separates the spectra of the two spin orientations: $+\mathpzc{h}$ corresponds to spin up and $-\mathpzc{h}$ to spin down. Thus, each spin channel hosts one bonding and one antibonding band, yielding four branches in total. In the absence of $\mathpzc{h}$, the up- and down-spin spectra overlap, recovering the spin-degenerate bonding-antibonding structure of the spinless Hatano-Nelson model.


\begin{thebibliography}{99}
\bibitem{pc1} I. O. Kulik, {\it Flux quantization in a normal metal}, JETP Lett. {\bf 11}, 275 (1970).

\bibitem{pc2} M. B\"{u}ttiker, Y. Imry, and R. Landauer, {\it Josephson behavior in small normal one-dimensional rings}, Phys. Lett. A {\bf 96}, 365 (1983).

\bibitem{pcex1} L. P. L\'{e}vy, G. Dolan, J. Dunsmuir, and H. Bouchiat, {\it Magnetization of mesoscopic copper rings: Evidence for persistent currents}, Phys. Rev. Lett. {\bf 64}, 2074 (1990).

\bibitem{pcex2} V. Chandrasekhar, R. A. Webb, M. J. Brady, M. B. Ketchen, W. J. Gallagher, and A. Kleinsasser, {\it Magnetic response of a single, isolated gold loop}, Phys. Rev. Lett. 67, 3578 (1991).

\bibitem{pc3} H. F. Cheung, Y. Gefen, E. K. Riedel, and W. H. Shih, {\it Persistent currents in small one-dimensional metal rings}, Phys. Rev. B {\bf 37}, 6050 (1988).

\bibitem{pc4} V. Ambegaokar and U. Eckern, {\it Coherence and persistent currents in mesoscopic rings}, Phys. Rev. Lett. {\bf 65}, 381 (1990).

\bibitem{pc5} B. L. Altshuler, Y. Gefen, and Y. Imry, {\it Persistent differences between canonical and grand canonical averages in mesoscopic ensembles: Large paramagnetic orbital susceptibilities}, Phys. Rev. Lett.
{\bf 66}, 88 (1991).

\bibitem{pc6} G. J. Jin, Z. D. Wang, A. Hu, and S. S. Jiang. {\it Persistent currents in mesoscopic Fibonacci rings}, Phys. Rev. B {\bf 55}, 9302 (1997).

\bibitem{pc7} V. Ferrari, G. Chiappe, E. V. Anda, and M. A. Davidovich, {\it Kondo Resonance Effect on Persistent Currents through a Quantum Dot in a Mesoscopic Ring}, Phys. Rev. Lett. {\bf 82}, 25 (1999).

\bibitem{pc8} J. Splettstoesser, M. Governale, and U. Z\"{u}licke, {\it Persistent current in ballistic mesoscopic rings with Rashba spin-orbit coupling}, Phys. Rev. B {\bf 68}, 165341 (2003).

\bibitem{pc9} S. K. Maiti, J. Chowdhury, and S. N. Karmakar, {\it Enhancement of persistent current in mesoscopic rings and cylinders: shortest and next possible shortest higher-order hopping}, J. Phys.: Condens. Matter {\bf 18}, 5349 (2006).

\bibitem{pc10} J. S. Sheng and K. Chang, {\it Spin states and persistent currents in mesoscopic rings: Spin-orbit interactions}, Phys. Rev. B {\bf 74}, 235315 (2006).

\bibitem{pc11} L. K. Castelano, G.-Q. Hai, B. Partoens, and F. M. Peeters, {\it Control of the persistent currents in two interacting quantum rings through the Coulomb interaction and interring tunneling}, Phys. Rev. B {\bf 78}, 195315 (2008).

\bibitem{pcex3} E. M. Q. Jariwala, P. Mohanty, M. B. Ketchen, and R. A. Webb, {\it Diamagnetic Persistent Current in Diffusive Normal-Metal Rings}, Phys. Rev. Lett. {\bf 86}, 1594 (2001).

\bibitem{pcex4} A. Fuhrerl, S. L\"{u}scherl, T. Ihn, T. Heinzel, K. Ensslin, W. Wegscheider, and M. Bichler, {\it Energy spectra of quantum rings}, Nature (London) {\bf 413}, 822 (2001).
(2001).

\bibitem{pcex5} R. Deblock, R. Bel, B. Reulet, H. Bouchiat, and D. Mailly, {\it Diamagnetic Orbital Response of Mesoscopic Silver Rings}, Phys. Rev. Lett. {\bf 89}, 206803 (2002).

\bibitem{pcex6} M. A. Castellanos-Beltran, D. Q. Ngo, W. E. Shanks, A. B. Jayich, and J. G. E. Harris, {\it Measurement of the Full Distribution of Persistent Current in Normal-Metal Rings}, Phys. Rev. Lett. {\bf 110}, 156801 (2013).




\bibitem{spc90} D. Loss, P. Goldbart, and A. V. Balatsky, {\it Berry's phase and persistent charge and spin currents in textured mesoscopic rings}, Phys. Rev. Lett. {\bf 65}, 1655 (1990).

\bibitem{spc94} Y. Zhou, H. Han, and L. L. Xue, {\it Spin-orbit coupling in one-dimensional conducting rings}, Phys. Rev. B {\bf 49}, 14010 (1994).

\bibitem{spc95} S. Oh and C.-M. Ryu, {\it Persistent spin currents induced by the Aharonov-Casher effect in mesoscopic rings}, Phys. Rev. B {\bf 51}, 13441 (1995).

\bibitem{spc99} J. Nitta, F. E. Meijer, and H. Takayanagi, {\it The Spin-interference device}, Appl. Phys. Lett. {\bf 75}, 695 (1999).

\bibitem{spc03a} J. Splettstoesser, M. Governale, and U. Z\"{u}licke, {\it Persistent current in ballistic mesoscopic rings with Rashba spin-orbit coupling}, Phys. Rev. B {\bf 68}, 165341 (2003).

\bibitem{spc03b} F. Sch\"{u}tz, M. Kollar, and P. Kopietz, {\it Persistent Spin Currents in Mesoscopic Heisenberg Rings}, Phys. Rev. Lett. {\bf 91}, 017205 (2003).

\bibitem{spc05} J.-N. Wu, M.-C. Chang, and M.-F. Yang, {\it Persistent spin current in mesoscopic ferrimagnetic spin ring}, Phys. Rev. B {\bf 72}, 172405 (2005).

\bibitem{spc07} Q.-F. Sun, X. C. Xie, and J. Wang, {\it Persistent Spin Current in a Mesoscopic Hybrid Ring with Spin-Orbit Coupling}, Phys. Rev. Lett. {\bf 98}, 196801 (2007).

\bibitem{spc08} Q.-F. Sun, X. C. Xie, and J. Wang, {\it Persistent spin current in nanodevices and definition of the spin current}, Phys. Rev. B {\bf 77}, 035327 (2008).

\bibitem{spcding} G.-H. Ding and B. Dong, {\it Spin-orbit coupling effect on persistent currents in a one-dimensional quantum ring with an Anderson impurity}, Phys. Rev. B {\bf 76}, 125301 (2007).

\bibitem{spc14} M. Ellner, N. Bol\'{i}var, B. Berche, and E. Medina, {\it Charge- and spin-polarized currents in mesoscopic rings with Rashba spin-orbit interactions coupled to an electron reservoir}, Phys. Rev. B {\bf 90}, 085305 (2014).

\bibitem{maiti14} S. K. Maiti, M. Dey, and S.N. Karmakar, {\it Persistent charge and spin currents in a quantum ring using Green's function technique: Interplay between magnetic flux and spin-orbit interactions}, Physica E: Low-dimensional Systems and Nanostructures {\bf 64}, 169 (2014).

\bibitem{maiti16} M. Patra and S. K. Maiti, {\it Characteristics of persistent spin current components in a quasi-periodic Fibonacci ring with spin-orbit interactions: Prediction of spin-orbit coupling and on-site energy}, Annals of Physics {\bf 375}, 337 (2016).

\bibitem{spc21} G. M. Maksimova, E. S. Gorlachev, A. V. Telezhnikov, and L. E. Golub, {\it Spin dynamics and equilibrium spin current in Aharonov-Casher rings with inhomogeneous Rashba spin-orbit interaction}, Semiconductors {\bf 55}, 764 (2021).

\bibitem{spc23} Y.-H. Hsu and Y.-P. Chen, {\it Spin current and internal Zeeman field in spin-orbit-coupled rings}, Phys. Rev. B {\bf 107}, 125419 (2023).

\bibitem{hatano1} N. Hatano and D. R. Nelson, {\it Localization Transitions in Non-Hermitian Quantum Mechanics}, Phys. Rev. Lett. {\bf 77}, 570 (1996).

\bibitem{hatano2} N. Hatano and D. R. Nelson, {\it Vortex pinning and non-Hermitian quantum mechanics}, Phys. Rev. B {\bf 56}, 8651 (1997).

\bibitem{nhreview} Y. Ashida, Z. Gong, and M. Ueda, {\it Non-Hermitian physics}, Adv. Phys. \textbf{69}, 249 (2020).

\bibitem{li2021} Q. Li, J.-J. Liu, and Y.-T. Zhang, {\it Non-Hermitian Aharonov-Bohm effect in the quantum ring}, Phys. Rev. B {\bf 103}, 035415 (2021).

\bibitem{ratchet2022} N. Treps, A. M. Corman, and C. S. Weiss, {\it Persistent current by a static non-Hermitian ratchet}, Phys. Rev. A 105, 023328 (2022).

\bibitem{shen2024} P.-X. Shen, Z. Lu, J. L. Lado, and M. Trif, {\it Non-Hermitian Fermi-Dirac Distribution in Persistent Current Transport}, Phys. Rev. Lett. {\bf 133}, 086301 (2024).

\bibitem{ganguly2025} S. Ganguly and S. K. Maiti, {\it Persistent current in a non-Hermitian Hatano-Nelson ring: Disorder-induced amplification}, Phys. Rev. B {\bf 111}, 195418 (2025).

\bibitem{ghosh2025} S. Ghosh, P. K. Ghosh, S. Sil, {\it Edge states and persistent current in a $\mathcal{PT}$-symmetric extended Su-Schrieffer-Heeger model with generic boundary conditions}, Phys. Rev. B {\bf 111}, 245428 (2025). 

\bibitem{sarkar2025} S. Sarkar, S. Satpathi, and S. K. Pati, {\it Enhancement of persistent current in a non-Hermitian disordered ring}, Phys. Rev. B {\bf 112}, 035425 (2025).

\bibitem{byers1961} N. Byers, and C. N. Yang, {\it Theoretical Considerations Concerning Quantized Magnetic Flux in Superconducting Cylinders}, Phys. Rev. Lett. {\bf 7}, 46 (1961).

\bibitem{biortho} D. C. Brody, {\it Biorthogonal quantum mechanics}, J. Phys. A: Math. Theor. {\bf 47}, 035305 (2013).


\bibitem{kawabata} K. Kawabata, K. Shiozaki, M. Ueda, and M. Sato, {\it Symmetry and topology in non-Hermitian physics}, Phys. Rev. X {\bf 9}, 041015 (2019).

\bibitem{shen2018} H. Shen, B. Zhen, and L. Fu, {\it Topological Band Theory for Non-Hermitian Hamiltonians}, Phys. Rev. Lett. {\bf 120}, 146402 (2018).

\bibitem{longhi2019} S. Longhi, {\it Metal-insulator phase transition in a non-Hermitian Aubry-André-Harper model}, Phys. Rev. B {\bf 100}, 125157 (2019).


\bibitem{moumita-prb} M. Dey, S. Sarkar, and S. K. Maiti, {\it Light irradiation controlled spin selectivity in a magnetic helix}, Phys. Rev. B {\bf 108}, 155408 (2023).

\bibitem{sarkar2019} S. Sarkar and S. K. Maiti, {\it Spin-selective transmission through a single-stranded magnetic helix}, Phys. Rev. B {\bf 100}, 205402 (2019).

\bibitem{yhsu} Y.-H. Su, S.-H. Chen, C. D. Hu, and C.-R. Chang, {\it Competition between spin–orbit interaction and exchange coupling within a honeycomb lattice ribbon}, J. Phys. D: Appl. Phys. {\bf 49}, 015305 (2015).

\bibitem{biortho1} L. Kohaupt, {\it Introduction to a Gram-Schmidt-type biorthogonalization method}, Rocky Mountain J. Math. {\bf 44}, 1265 (2014). 

\bibitem{biortho2} P. Zhong, W. Pan, H. Lin, X. Wang, and S. Hu, {\it Density matrix renormalization group algorithm for non-Hermitian systems}, arXiv:2401.15000v3, Phys. Rev. Lett. (to be published).

\bibitem{anderson} P. W. Anderson, {\it Absence of Diffusion in Certain Random Lattices}, Phys. Rev. {\bf 109}, 1492 (1958).

\bibitem{gangof4} E. Abrahams, P. W. Anderson, D. C. Licciardello, and T. V. Ramakrishnan, {\it Scaling Theory of Localization: Absence of Quantum Diffusion in Two Dimensions}, Phys. Rev. Lett. {\bf 42}, 673 (1979).

\bibitem{cc-wan} C. C. Wanjura1, M. Brunelli, and A. Nunnenkamp, {\it Correspondence between Non-Hermitian Topology and Directional Amplification in the Presence of Disorder}, Phys. Rev. Lett. {\bf 127}, 213601 (2021).

\bibitem{longi} S. Longhi, {\it Phase transitions in a non-Hermitian Aubry-Andr\'{e}-Harper model}, Phys. Rev. B {\bf 103}, 054203 (2021).

\bibitem{aah1} P. G. Harper, {\it The general motion of conduction electrons
in a uniform magnetic field, with application to the diamagnetism of metals}, Proc. R. Soc. London, Ser. A {\bf 68},
874 (1955).

\bibitem{aah2} S. Aubry and G. Andr\'{e}, {\it Analyticity breaking and Anderson localization in incommensurate lattices}, Ann. Isr. Phys. Soc. {\bf 3}, 133 (1980).

\bibitem{aah3} S. Sil, S. K. Maiti, and A. Chakrabarti, {\it Metal-Insulator Transition in an Aperiodic Ladder Network: An Exact Result}, Phys. Rev. Lett. {\bf 101}, 076803 (2008).










\end{thebibliography}
\end{document}